%% file: main.tex
\documentclass[11pt]{article}

\usepackage{amsthm}
\usepackage{graphicx} 
\usepackage{array} 

\usepackage{amsmath, amssymb, amsfonts, verbatim}
\usepackage{hyphenat,epsfig,subcaption,multirow}
\usepackage{nicefrac}
\usepackage{paralist}
\usepackage[font=small,labelfont=bf]{caption}

\usepackage[table, x11names, usenames,dvipsnames]{xcolor}
\usepackage[ruled]{algorithm2e}

\DeclareFontFamily{U}{mathx}{\hyphenchar\font45}
\DeclareFontShape{U}{mathx}{m}{n}{
      <5> <6> <7> <8> <9> <10>
      <10.95> <12> <14.4> <17.28> <20.74> <24.88>
      mathx10
      }{}
\DeclareSymbolFont{mathx}{U}{mathx}{m}{n}
\DeclareMathSymbol{\bigtimes}{1}{mathx}{"91}

\usepackage{tcolorbox}
\tcbuselibrary{skins,breakable}
\tcbset{enhanced jigsaw}

\usepackage[normalem]{ulem}
\usepackage[compact]{titlesec}

\definecolor{DarkRed}{rgb}{0.5,0.1,0.1}
\definecolor{DarkBlue}{rgb}{0.1,0.1,0.5}

\usepackage{nameref}
\definecolor{ForestGreen}{rgb}{0.1333,0.5451,0.1333}
\definecolor{Red}{rgb}{0.9,0,0}
\usepackage[linktocpage=true,
	pagebackref=true,colorlinks,
	linkcolor=DarkRed,citecolor=ForestGreen,
	bookmarks,bookmarksopen,bookmarksnumbered]
	{hyperref}
\usepackage[noabbrev,nameinlink]{cleveref}
\crefname{property}{property}{Property}
\creflabelformat{property}{(#1)#2#3}
\crefname{equation}{eq}{Eq}
\creflabelformat{equation}{(#1)#2#3}

\usepackage{bm}
\usepackage{url}
\usepackage{xspace}
\usepackage[mathscr]{euscript}

\usepackage{tikz}
\usetikzlibrary{arrows}
\usetikzlibrary{arrows.meta}
\usetikzlibrary{shapes}
\usetikzlibrary{backgrounds}
\usetikzlibrary{positioning}
\usetikzlibrary{decorations.markings}
\usetikzlibrary{patterns}
\usetikzlibrary{calc}
\usetikzlibrary{fit}
\usetikzlibrary{decorations}

\usepackage{mdframed}

\usepackage[noend]{algpseudocode}
\makeatletter
\def\BState{\State\hskip-\ALG@thistlm}
\makeatother

\usepackage{cite}
\usepackage{enumitem}

\usepackage[margin=1in]{geometry}

\newtheorem{theorem}{Theorem}
\newtheorem{lemma}{Lemma}[section]
\newtheorem{proposition}[lemma]{Proposition}

\newtheorem{claim}[lemma]{Claim}

\newtheorem{definition}[lemma]{Definition}

\newtheorem*{claim*}{Claim}
\newtheorem*{assumption*}{Assumption}
\newtheorem*{proposition*}{Proposition}
\newtheorem*{lemma*}{Lemma}
\newtheorem*{problem5*}{Problem}

\crefname{lemma}{Lemma}{Lemmas}
\crefname{claim}{claim}{claims}

\newtheorem{mdresult}[theorem]{Theorem}
\newenvironment{result}{\begin{mdframed}[backgroundcolor=lightgray!40,topline=false,rightline=false,leftline=false,bottomline=false,innertopmargin=2pt]\begin{mdresult}}{\end{mdresult}\end{mdframed}}

\theoremstyle{definition}

\newtheorem{observation}[lemma]{Observation}

\newtheoremstyle{restate}{}{}{\itshape}{}{\bfseries}{~(restated).}{.5em}{\thmnote{#3}}
\theoremstyle{restate}

\allowdisplaybreaks

\renewcommand{\leq}{\leqslant}
\renewcommand{\geq}{\geqslant}

\renewcommand{\qed}{\nobreak \ifvmode \relax \else
      \ifdim\lastskip<1.5em \hskip-\lastskip
      \hskip1.5em plus0em minus0.5em \fi \nobreak
      \vrule height0.75em width0.5em depth0.25em\fi}

\newcommand{\Qed}[1]{\rlap{\qed$_{\textnormal{~\Cref{#1}}}$}}

\input{macros}

\title{Settling the Pass Complexity of Streaming Set Cover\footnote{A preliminary version of this paper appeared in the ACM Symposium on Theory of Computing, STOC 2026.}
}

\author{
	Sepehr Assadi\footnote{(sepehr@assadi.info) School of Computer Science, University of Waterloo. 
		Supported in part by an NSERC
	Discovery Grant and a Faculty of Math Research Chair grant from University of Waterloo. \smallskip} 
	\\ {\small University of Waterloo} \and
	Janani Sundaresan\footnote{(js16503@nyu.edu) Courant Institute of Mathematical Sciences, New York University. This work was done when the author was a graduate student in the University of Waterloo. \smallskip} 
	\\ {\small New York University}
}

\date{}

\begin{document}
\maketitle

\pagenumbering{roman}

\input{abstract}
\clearpage



\bigskip

\setcounter{tocdepth}{3}
\tableofcontents

\clearpage

\pagenumbering{arabic}
\setcounter{page}{1}

\input{intro}

\input{prelim}

\input{proof}

\bibliographystyle{halpha-abbrv}
\bibliography{new}

\input{figfull}

\clearpage

\appendix
\part*{Appendix}

\input{thealg}

\end{document}

%% file: macros.tex
\newcommand{\Ot}{\ensuremath{\widetilde{O}}}

\newcommand{\Paren}[1]{\Big(#1\Big)}

\newcommand{\bracket}[1]{\left[#1\right]}
\newcommand{\paren}[1]{\ensuremath{\left(#1\right)}\xspace}
\newcommand{\card}[1]{\left\vert{#1}\right\vert}
\newcommand{\Omgt}{\ensuremath{\widetilde{\Omega}}}

\newcommand{\IN}{\ensuremath{\mathbb{N}}}

\newcommand{\expect}[1]{\Exp\bracket{#1}}

\newcommand{\set}[1]{\ensuremath{\left\{ #1 \right\}}}
\newcommand{\poly}{\mbox{\rm poly}}

\newcommand{\opt}{\textnormal{\ensuremath{\mbox{opt}}}\xspace}

\newcommand{\alg}{\ensuremath{\mathcal{A}}\xspace}

\DeclareMathOperator*{\Exp}{\ensuremath{{\mathbb{E}}}}
\DeclareMathOperator*{\Prob}{\ensuremath{\textnormal{Pr}}}
\renewcommand{\Pr}{\Prob}

\newenvironment{tbox}{\begin{tcolorbox}[
		enlarge top by=5pt,
		enlarge bottom by=5pt,
		 breakable,
		 boxsep=0pt,
                  left=4pt,
                  right=4pt,
                  top=10pt,
                  arc=0pt,
                  boxrule=1pt,toprule=1pt,
                  colback=white
                  ]
	}
{\end{tcolorbox}}

\newcommand{\II}{\ensuremath{\mathbb{I}}}

\newcommand{\mireal}[1][]{
  \ifx\relax#1\relax%
    \II(\mione \,; \mitwo)%
  \else%
    \II(\mione \,; \mitwo\mid #1)%
  \fi
}

\newcommand{\cU}{\ensuremath{\mathcal{U}}}
\newcommand{\cS}{\ensuremath{\mathcal{S}}}
\newcommand{\cC}{\ensuremath{\mathcal{C}}}

\newcommand{\PC}[1]{\ensuremath{\textnormal{\textsf{PC}}_{#1}}}

\newcommand{\prot}{\ensuremath{\pi}}
\newcommand{\player}[1]{\ensuremath{P^{#1}}}
\newcommand{\func}[1]{\ensuremath{f^{#1}}}
\newcommand{\Sset}[2]{\ensuremath{S}^{#1}_{#2}}
\newcommand{\Sstar}[1]{\ensuremath{S^{#1}_{\star}}}

\newcommand{\Uset}[2]{\ensuremath{U^{#1}_{#2}}}

\newcommand{\jstar}{\ensuremath{j^{\star}}}

%% file: abstract.tex
\begin{abstract}

\medskip

In the streaming set cover problem, $m$ sets from a universe of size $n$ are arriving one by one in a stream, and the algorithm 
is allowed to process the stream using one or a few passes and a space of $o(mn)$, which is sublinear in the input size. The goal is to 
determine the minimal (or approximately minimal) number of sets that cover the universe at the end of the last pass. 

\medskip

This problem has been studied extensively over the years with rapid progress that led to several $O(\log{n})$-approximation algorithms in $\Ot(mn^{1/p})$ space and $O(p)$ passes. 
However, progress on this front has largely stagnated over the past decade, despite the absence of any lower bounds that rule out even an $O(\log{n})$-approximation in $O(m)$ space and just two passes. 

\medskip
 
We provide a simple explanation for this lack of progress by establishing an \emph{optimal} three-way \textbf{space-pass-approximation} tradeoff for this problem: any $\alpha$-approximation algorithm for streaming set cover requires
\[
		\Omgt\Paren{\frac{m}{\alpha} \cdot \paren{\frac{n}{\alpha}}^{1/p}}
\]
space in $p$ passes whenever $\alpha \ll n^{1/(p+1)}$. 

\medskip
In light of prior work, this result is optimal up to constant factors in $p$ and logarithmic factors in $n,m$ for any $\alpha\geq p$. Our bound is optimal with respect to the range of $\alpha$ also, and fully settles the complexity
of this fundamental problem in the streaming model. The proof of this result is (surprisingly) simple and non-technical and relies on a randomized reduction from a variant of the standard pointer chasing problem in communication complexity, using elementary properties of random sets.

\end{abstract}

%% file: intro.tex

\newcommand{\Omegat}{\tilde{\Omega}}
\newcommand{\estar}{e_\star}
\newcommand{\itemref}[1]{\hyperref[#1]{Item-\ref{#1}}}
 
\section{Introduction}

Set cover is a standard optimization problem with a wide range of applications \cite{GrossmanW97,Mihail99,GolabKKSY08,SahaG09,ChierichettiKT10}. There is a universe $\cU$ of size $n$, and there is a family $\cS$ containing $m$ subsets $S_1, \ldots, S_m \subseteq \cU$. The aim is to find a collection $\cC \subseteq \cS$ such that the union of all the subsets in collection $\cC$ covers the entire universe $\cU$, while minimizing the size of the collection $\cC$.  There exist multiple algorithms that achieve an approximation ratio of $\ln n$ \cite{Chvatal79,Johnson74,Lovasz75}, and we also know that this ratio is the best possible in polynomial time if P $\neq $ NP \cite{Feige98,DinurS14}.

In recent years, there has also been extensive interest in studying set cover in the streaming model. Here, the sets arrive one by one in some arbitrary order in a stream. The algorithm is allowed to make one or few passes over the stream before outputting a solution. Storing all the sets explicitly to process the set cover instance needs $\Theta(mn)$ space, but the goal is to find a solution in $o(mn)$ space. As is standard in the streaming model, we primarily focus 
on the space complexity of the algorithms and do not necessarily bound their runtime. The study of set cover in streaming was initiated by \cite{SahaG09}, and since then has been extensively explored in many works \cite{DemaineIMV14,EmekR16,ChakrabartiW16,HarPeledIMV16,AssadiKL16,Assadi17,IndykMRUVY17,BateniEV17,KhannaK22,KhannaKA23}. 

The work in this area can be broadly classified as having two main focuses: using as little space as possible, or getting as good an approximation as possible. 

 \paragraph{Low Space.} The goal in the low-space regime is to use $\Ot(n)$ space\footnote{Throughout this paper,  $\widetilde{\Omega}$ and $\widetilde{O}$ hide logarithmic factors in $m, n$.} and minimize  approximation ratio and number of passes as much as possible.\footnote{Space of $O(n)$ is necessary to even certify whether the input set cover instance has any feasible solution.} \cite{SahaG09} initiated this line of study and obtained an $O(\log n)$-approximation in $O(\log n)$-passes.~\cite{EmekR16} considered single-pass algorithms and showed a tight bound of $\Theta(\sqrt{n})$ on the approximation ratio. Finally, \cite{ChakrabartiW16} showed a smooth trade-off between the results of \cite{EmekR16} and \cite{SahaG09}: for any $p \geq 1$, they gave an algorithm that obtains an $\Ot(n^{1/(p+1)})$-approximation in $p$-passes and showed that obtaining $\tilde{\Omega}(n^{1/(p+1)})$ approximation requires $\Omega(n^c)$ space for some constant $c > 1$, fully settling the 
 complexity of the streaming set cover problem in the low space regime. 
	
 \paragraph{Low Approximation.} Another line of work---which is the main focus of our work---aims to match the performance of non-streaming algorithms in terms of approximation ratio 
 while using sublinear space, although not necessarily as restrictive as $\Ot(n)$ space. Typically, these results either use polynomial time to get $O(\log n)$-approximations or exponential time to get $O(1)$-approximations. 
 Here, we often assume that $m$ is some (large) polynomial in $n$. 
 
 The first result in this regime---although not studied directly for streaming algorithms---is the work of~\cite{Nisan02} who proved that communication complexity of getting a better than a $(\log{n}/2)$-approximation
 to set cover is $\Omega(m)$ bits. This translates to a lower bound of $\Omega(m/p)$ space for $p$-pass algorithms and same approximation ratio using the standard connection between streaming algorithms and communication protocols
 (see, e.g.~\cite{AlonMS96}).  

Algorithmically, \cite{DemaineIMV14} initiated this line of study and obtained an $O(p)$-approximation in $p$-passes and $\Ot(mn^{1/O(\log(p))})$-space. This space bound was later improved by \cite{HarPeledIMV16} to $\Ot(mn^{O(1/p)})$ space. 
For polynomial-time algorithms, this result led to an $O(p\log{n})$-approximation which was improved to $O(\log{n})$-approximation in~\cite{BateniEV17} and even $(1+o(1)) \cdot \ln{n}$ approximation in~\cite{IndykMRUVY17} 
at a cost of increasing the number of passes by a constant factor. This constitutes the state-of-the-art algorithms for this problem. 

The tradeoff curve obtained by algorithms of~\cite{HarPeledIMV16,BateniEV17,IndykMRUVY17} namely, $mn^{1/p}$ space and $O(p)$ passes, is a familiar one in the streaming model. Based on this,~\cite{HarPeledIMV16} conjectured:

\begin{quote}
 \emph{``... this tradeoff might be tight, at least for `low enough' approximation factors''}. 
\end{quote}

In the same work,~\cite{HarPeledIMV16} provided an evidence in favor of this conjecture: finding an \emph{exact} set cover, i.e., a $1$-approximation, requires $\Omegat(mn^{1/2p})$ space in $p$-passes. 
It turns out however that this evidence is not that strong due to the \emph{three-way} interplay between the resource parameters of the problem, space, pass, and approximation. Specifically,~\cite{Assadi17} subsequently showed 
that finding an exact set cover in $p$ passes requires $\Omega(mn/p)$ space---and not just $\Omega(mn^{1/p})$ proven by~\cite{HarPeledIMV16}---and more generally $\Omega(mn^{1/\alpha}/p)$ space for $\alpha$-approximation when $\alpha = o(\log{n})$. 
The bounds of~\cite{Assadi17} identifies the optimal \emph{space-approximation} tradeoff for this problem when we \emph{ignore} the number of passes (as long as it is $\poly\!\log\!{(n)}$); but it does not address the original conjecture (only weakens the evidence of~\cite{HarPeledIMV16}). 

Focusing on the original goal of low space algorithms---matching the performance of non-streaming algorithms, namely, a $\log{n}$ or even a $\poly\!\log\!{(n)}$-approximation---and the conjecture of~\cite{HarPeledIMV16}, 
we do not yet have any conclusive answer. On one hand, the algorithmic progress on this question has halted for nearly a decade now. On the other hand, even considering all known lower bounds~\cite{Nisan02,EmekR16,DemaineIMV14,AssadiKL16,ChakrabartiW16,HarPeledIMV16,Assadi17}, it is still entirely plausible that there is even a two-pass algorithm with $O(m)$ space and $O(\log{n})$-approximation! 

\subsection{Our Contributions}

We settle the conjecture of~\cite{HarPeledIMV16} by proving that any $\log{n}$ or even $\poly\!\log\!{(n)}$-approximation algorithm for streaming set cover 
requires $\Omgt(mn^{1/p})$ space in $p$-passes. In fact, we prove a more general result that establishes a (nearly) \emph{optimal} three-way space-pass-approximation tradeoff:

\begin{result}\label{res:main}
	For any $p, \alpha, m, n \geq 1$, any $p$-pass algorithm that given a set cover instance over universe $[n]$ with $m$ input sets outputs an $\alpha$-approximation to the optimal set cover \emph{size} (not necessarily even finding the sets)
	 with probability at least $2/3$ requires space 
	 \[
	\tilde{\Omega}\paren{\frac{m}{\alpha} \cdot \paren{\frac{n}{\alpha}}^{1/p}}.
	\]
	The lower bound holds as long as $p \ll \dfrac{\log{n}}{\log\log{n}}$ and $p \leq \alpha  \ll \paren{\dfrac{1}{p} \cdot (\dfrac{n}{\log{m}})^{1/(p+1)}}$. 
\end{result}
Several remarks are in order (see also~\Cref{tab:results1,tab:results} and~\Cref{fig:tradeoff} before the Appendix), 
\vspace{-2pt}
\begin{itemize}[leftmargin=10pt, itemsep=1pt]
	\item The tradeoff in~\Cref{res:main} is nearly optimal up to constant factors in $p$ for all parameters in light of existing results by a simple modification of the algorithms in~\cite{HarPeledIMV16,AssadiKL16} (see~\Cref{sec:setcov-ub}). 
	\item The ranges of parameters in~\Cref{res:main} are all necessary and nearly optimal. For any $p = \Omega(\log{n})$, we can obtain an $O(\log{n})$ approximation in $\Ot(n)$ space and $O(\log{n})$ passes~\cite{SahaG09,ChakrabartiW16}. When $\alpha < p$, the space requirement ``jumps'' to $\Omega(mn^{1/\alpha}/p)$ by the lower bound of~\cite{Assadi17}. When $\alpha \geq n^{1/(p+1)}$, 
	the space requirement ``drops'' to $\Ot(n)$ by the result of~\cite{ChakrabartiW16}. 
	\item \Cref{res:main} generalizes the \emph{single-pass} lower bounds of~\cite{AssadiKL16} who proved an $\Omega(mn/\alpha^2)$ space lower bound (matching our tradeoff for $p=1$) to all choices of $p \ll \log{n}$. 
	\item \Cref{res:main} also generalizes the \emph{multi-pass} lower bounds of~\cite{ChakrabartiW16} by proving that $\Omega(m)$ space is needed for $(n/\log{n})^{1/(p+1)}$ approximation, 
	improving upon their $\Omega(m^{1/p})$ space lower bound (at a cost of a $(\log{n})^{1/(p+1)}$ loss in our approximation ratio). 
\end{itemize}

\input{table}

\subsection{Our Techniques}

The closest work to us is the previous streaming set cover lower bound of~\cite{ChakrabartiW16} in the low space regime. 
To prove a $p$-pass lower bound, they start with ``many'' $(p-1)$-pass hard instances on different universes with a ``small'' overlap between them. 
The stream consists of these sets plus a single set at the end that covers all elements except for the universe of one randomly chosen \emph{special} $(p-1)$-pass hard instance. 
The small overlap between the sets implies that getting a good approximation to the entire instance requires one to solve the special sub-instance also. The space lower bound is proven
by arguing that the number of embedded $(p-1)$-pass instances is so much larger than the memory of the algorithm, that it is unlikely in the first pass, the algorithm remembers anything useful about the special one; thus, 
it needs to solve a hard $(p-1)$-pass instance in the remaining $(p-1)$ passes which, by induction, is not possible. This argument is formalized in~\cite{ChakrabartiW16} via a reduction
from the \emph{tree pointer jumping} problem in communication complexity~\cite{ChakrabartiCM08} and introducing an extremal family of small overlapping sets, named \emph{edifices}, through well-structured algebraic geometry constructs \cite{Buekenhout79}. 

The arguments in~\cite{ChakrabartiW16} are heavily tailored to the low space regime: as argued above,  the \emph{number} of underlying $(p-1)$-pass instances needs to be more than the memory of the algorithm, 
which leads to an exponential growth in the number of inputs sets across the passes. In terms of $m$, this only leads to an $\Omega(m^{1/p})$ space lower bound, which is way below the $\omega(m)$ bounds one needs in the low approximation regime. 
Recently, there have been successful attempts in extending similar arguments to prove graph streaming lower bounds\footnote{Streaming set cover is equivalent to dominating set in graph streaming.~\cite{ChakrabartiW16} gives an $\Omega(n^{1/p})$ space lower bound for $n$-vertex graphs whereas one ideally would like an $\Omega(n^{1+1/p})$-space lower bound in graph streams.}  using the newly developed \emph{hierarchical embedding} technique~\cite{AssadiKNS24,AssadiBKNS25}. 
However, the reliance of this technique---similar to~\cite{ChakrabartiW16}---in using \emph{separate} $(p-1)$-pass hard sub-instances (in terms of the sets in those instances), make it similarly inapplicable to proving~\Cref{res:main}.  

\paragraph{Our approach.} We take a different approach when proving~\Cref{res:main}. For simplicity, let us focus on the case when $\alpha \approx \log{n}$. Our $p$-pass hard instances are also constructed 
recursively from $(p-1)$-pass hard instances, but not so directly as in~\cite{ChakrabartiW16}. 
A $p$-pass hard instance consists of a \emph{single} $(p-1)$-pass hard instance with an \emph{unknown} universe; instead, it will be revealed to the algorithm at the end of the first pass---by our construction of the instance---%
as essentially a \emph{random subset} of size $\approx n^{1-1/p}$ from the original universe (once the universe is revealed, 
the $(p-1)$-pass instance simply becomes the intersection of the original sets with this new universe). 

Unlike in~\cite{ChakrabartiW16} (and~\cite{AssadiKNS24,AssadiBKNS25}), there is no hope in proving that the algorithm does not learn \emph{any} information about the underlying $(p-1)$-pass hard instance (given it is just one instance). 
But, we show that the algorithm does not learn anything \emph{useful} about this underlying instance. This is achieved through a careful construction of the input
instances as sets of size $\approx n/\log{n}$ that are \emph{marginally} uniformly random, but have a mild \emph{correlation} in a way that each set ``points to'' another one; the union
of these two sets covers (considerably) more elements than two independent random sets. This correlation is then used to argue that to get a good approximation, we should start from some set---whose identity is forced to us at the end of the stream---%
and then continue picking the sets ``pointed to'' by the previous one. At the same time,  this correlation is low enough that until the algorithm has not identified the universe of the underlying $(p-1)$-hard instance, 
its chance of revealing useful information about the \emph{right} set of pointers is negligible\footnote{To compare this with~\cite{ChakrabartiW16}, for them one only needs to reveal $\log{n}$ bits about a $(p-1)$-pass instance 
in the first pass to make it no longer hard in the remaining $(p-1)$ passes; but there are so many such instances that the algorithm cannot reveal $\log{n}$ bits for each of them. On the other hand, 
to break our underlying $(p-1)$-pass instance \emph{before} its universe of size $n^{1-1/p}$ is revealed, one still needs to reveal $\Omega(mn^{1/p})$ information.}.

To formalize our result, we use the existing lower bounds for the \emph{pointer chasing} problem introduced by\cite{PapadimitriouS84}, which has since became a staple in proving multi-pass streaming lower bounds of similar nature, e.g., in~\cite{GuruswamiO13,AssadiKSY20,ChenKPSSY21,AssadiGLMM24}. Our reduction from pointer chasing is fairly direct and we rely on elementary properties of random sets to carry out the analysis. 

We conclude this part by remarking that the dynamics of our hard instances, namely, reducing the problem to another instance on $\approx n^{1-1/p}$ elements after the first pass, 
is exactly what streaming set cover algorithms in~\cite{HarPeledIMV16,BateniEV17,Assadi17,IndykMRUVY17} achieve through their \emph{element sampling technique}. 
Indeed, our lower bound construction is directly inspired by these algorithms.

%% file: table.tex


\def\arraystretch{1.7}
\begin{table}[h!]
\centering
{
\begin{tabular}{|c | c | c | c | c |}
 \hline
  ~~~~Citation~~~~ & ~~Space~~ & ~~~~Passes~~~~ & ~~Approximation~~ & ~~~~~~Range of Parameters~~~~~~\\ 
  \hline\hline
  \rowcolor{gray!20}  \multicolumn{5}{|c|}{Low Space Regime} \\
  \hline\hline
  \cite{EmekR16} & $\Omega(m)$ & $1$ & $\sqrt{n}$  & \\
  \hline\hline
  \cite{ChakrabartiW16} & $\Omgt(m^{1/p})$ & $p$ &  $n^{1/(p+1)}$ & $p \ll \frac{\log{n}}{\log\log{n}}$  \\
  \hline\hline
  \rowcolor{gray!20}  \multicolumn{5}{|c|}{Low Approximation Regime} \\
  \hline\hline
  \cite{AssadiKL16} & $\Omega(mn/\alpha^2)$ & $1$ & $\alpha$  & $\alpha \ll \sqrt{n}$ \\
  \hline\hline
  \cite{Nisan02} & $\Omega(m/p)$ & $p$ & $(\log{n})/2$  & \\
  \hline\hline
  \cite{HarPeledIMV16} & $\Omega(mn^{1/p})$ & $p$ & $1$  & $p \ll \frac{\log{n}}{\log\log{n}}$ \\
  \hline\hline
  \cite{Assadi17} & $\Omega(mn^{1/\alpha}/p)$ & $p$ & $\alpha$  & $\alpha \ll \log{n}$ \\
  \hline\hline
\rowcolor{blue!10}   \textbf{This work} & $\Omgt(mn^{1/p})$ & $p$ & $\poly\!\log\!{(n)}$  & $p \ll \frac{\log{n}}{\log\log{n}}$  \\
  \hline\hline
\rowcolor{blue!10}   \textbf{This work} & $\Omgt\paren{\frac{m}{\alpha} \cdot (\frac{n}{\alpha}}^{1/p})$ & $p$ & $\alpha$  & $p \ll \frac{\log{n}}{\log\log{n}}$ , $p \leq \alpha \ll n^{1/(p+1)}$ \\
  \hline  
\end{tabular}
}
\caption{A summary of prior lower bounds for streaming set cover and our work.}\label{tab:results1}
\end{table} 

\medskip

\begin{table}[h!]
\centering
{
\begin{tabular}{|c | c | c | c | c |}
 \hline
  ~~~~Citation~~~~ & ~~Space~~ & ~~~~Passes~~~~ & ~~Approximation~~\\ 
  \hline\hline
  \cite{ChakrabartiW16} & $\Ot(n)$ & $p$ &  $n^{1/(p+1)}$   \\
   \hline\hline
   \cite{HarPeledIMV16} & $\Ot(mn^{1/p})$ & $O(p)$ & $p$   \\
  \hline\hline
  \cite{IndykMRUVY17} & $\Ot(mn^{1/p})$ & $O(p)$ & $\ln{n}$    \\
  \hline\hline
 \cite{HarPeledIMV16} +  \cite{AssadiKL16} & $\Ot\!\paren{\frac{m}{\alpha} \cdot (\frac{n}{\alpha}}^{1/p})$ & $O(p)$ & $\alpha$  \\
  \hline  
\end{tabular}
}
\caption{The state-of-the-art multi-pass algorithms for streaming set cover. The final line is not stated explicitly in prior work but follows 
from a combination of~\cite{HarPeledIMV16,AssadiKL16} (see~\Cref{sec:setcov-ub}) and proves the optimality of our lower bound. The upper bound on $\alpha$ in 
our lower bound is also optimal in light of the algorithm in the first line.
}\label{tab:results}
\end{table}

%% file: prelim.tex
\section{Preliminaries}

We use $[n]$ to denote the set $\set{1, 2 \ldots, n}$ for any $n \in \IN$. For the cartesian product of two sets $[c] \times [d]$, we use $(i,j)$ to denote the elements in the set for $i \in [c]$ and $j \in [d]$.

We use the following standard form of Chernoff bound and its extension to negatively correlated variables in~\cite{PanconesiS97}. 

\begin{proposition}[Chernoff Bound; cf.~\cite{DubhashiP09}]\label{prop:chernoff}
	Let $X_1,\ldots,X_n$ be $n$ independent random variables in $[0,1]$ and $X := \sum_{i=1}^{n} X_i$. For any $\delta > 0$ and $\mu_{min} \leq \expect{X} \leq \mu_{max}$, 
	\begin{align*}
		&\Pr\paren{X \geq (1+\delta) \cdot \mu_{max}} \leq \exp\paren{-\frac{\delta^2 \cdot \mu_{max}}{2+\delta}} \quad , \quad \Pr\paren{X \leq (1-\delta) \cdot \mu_{min}} \leq \exp\paren{-\frac{\delta^2 \cdot \mu_{min}}{2}}.
	\end{align*} 	
	Moreover, the upper tail bound continues to hold as long $X_1,\ldots,X_n$ are negatively correlated, i.e., for every $S \subseteq [n]$, 
	\[
	\expect{\prod_{i \in S} X_i} \leq \prod_{i \in S} \expect{X_i}. 
	\]
\end{proposition}

\paragraph{Pointer chasing.} We now formally define the pointer chasing problem; see also~\Cref{fig:pc}. This is the standard definition in the multi-party setting that has been used in multiple results in the past, e.g.,~\cite{GuruswamiO13,AssadiN21}, although many variants exist. 

\begin{definition}\label{def:pc}
	For any integers $p, a, b \geq 1$, the pointer chasing problem $\PC{p, a, b}$ is defined as:
	\begin{itemize}
		\item There are $p+1$ players $\player{1}, \player{2}, \ldots, \player{p+1}$. 
		\item $\player{1}$ has a balanced Boolean function $\func{1}:[a] \times [b] \rightarrow \{0,1\}$.\footnote{By balanced, we mean number of inputs mapping to 0 and number of inputs mapping to 1 are equal.}
		\item For each $ 1 < i \leq (p+1)$, $\player{i}$ has a function $\func{i}:[a] \times [b] \rightarrow [a] \times [b]$. 
		\item In one round, they talk in the order of $\player{1}, \player{2}, \ldots, \player{p+1}$ on a blackboard that everyone can see.  They talk for a total of $p$ rounds. 
		\item For each $i \in [p+1]$, we define the \textbf{special pointer} $z_i \in [a] \times [b]$ recursively as:
		\[
		z_{p+1} = (1,1),  \textnormal{ and for $1 \leq i < (p+1)$, }  z_{i} = \func{i+1}(z_{i+1}).
		\]
		\item At the end of all the rounds, the last player $\player{p+1}$ has to output the value of $\func{1}(z_1)$
		which is a Boolean value, based on their input and the contents of the blackboard. 
	\end{itemize}
	See \Cref{fig:pc} for an illustration. 
\end{definition}

We know the following lower bound on the communication complexity of $\PC{p, a, b}$.

\begin{proposition}[\!\!\cite{AssadiN21}]\label{prop:pc-lb}
	For any $p \geq 1$, any randomized protocol $\prot$ that solves $\PC{p, a, b}$ with probability at least $3/5$ has the total length of messages on the board to be at least \[
	\Omega\paren{\frac{a \cdot b}{p^5}}.
	\]   
\end{proposition}

The work of \cite{AssadiN21} proves a lower bound for pointer chasing when the functions given to each player are permutations, but this is a special case of our problem, and hence the lower bound holds for our version of pointer chasing also.

\begin{figure}[h!]
	\begin{center}
		\resizebox{0.6\textwidth}{7cm}{
		\input{tikz-pc}
	}
	\end{center}
	\caption{An illustration of pointer chasing for $p = 3, a = 3$ and $b = 4$. Each function is represented by a $3 \times 4$ grid containing the corresponding values. The special pointers of each player are highlighted in green. The red dashed arrows represent how the pointers move. The answer to the pointer chasing instance is $0$. }
	\label{fig:pc}
\end{figure}


%% file: tikz-pc.tex
\begin{tikzpicture}[line cap=round, line join=round]
	\pgfmathsetlengthmacro{\CW}{0.90cm}   
	\pgfmathsetlengthmacro{\CH}{0.75cm}   
	\pgfmathsetlengthmacro{\PAD}{0.15cm}  
	\pgfmathsetlengthmacro{\GX}{1.20cm}   
	\pgfmathsetlengthmacro{\GY}{1.80cm}   
	\pgfmathsetlengthmacro{\GW}{4*(\CW+\PAD)} 
	\pgfmathsetlengthmacro{\GH}{3*(\CH+\PAD)} 
	
	\pgfmathsetseed{42} 
	
	\begin{scope}
	\draw[rounded corners=3pt, thick, opacity=0.4]
	(-0.3cm, -0.3cm) rectangle (\GW+0.3cm, \GH+1.1cm);
	
	\node[font=\bfseries\large] at (.5*\GW, \GH + 0.6cm) {$P^1$};
	
	\foreach \r in {0,...,2}{%
		\foreach \c in {0,...,3}{%
			\pgfmathsetlengthmacro{\XX}{\c*(\CW+\PAD)}
			\pgfmathsetlengthmacro{\YY}{\r*(\CH+\PAD)}
			\coordinate (P1-r\r c\c) at (\XX+\CW, \YY+0.5*\CW);
			\coordinate (P1-r1\r c1\c) at (\XX+\CW+0.2*\CW, \YY+0.5*\CW);
			\coordinate (P1-r2\r c2\c) at (\XX+1.2*\CW, \YY-\CW);
			
			\ifnum \r = 2
				\ifnum \c = 3
					\path[draw, fill=ForestGreen!15] (\XX,\YY) rectangle ++(\CW,\CH); 
				\else
				\path[draw, fill=gray!5] (\XX,\YY) rectangle ++(\CW,\CH); 
				\fi
			\else
			\path[draw, fill=gray!5] (\XX,\YY) rectangle ++(\CW,\CH); 
			\fi
			\pgfmathsetlengthmacro{\XC}{\XX + 0.5*\CW}
			\pgfmathsetlengthmacro{\YC}{\YY + 0.5*\CH}
			\pgfmathrandominteger{\v}{0}{1}
			\ifnum \r=1
				\ifnum \c=1
				\node at (\XC,\YC) {0};
				\else
				\node at (\XC,\YC) {\v};
				\fi
			\else
				\node at (\XC,\YC) {\v};
			\fi
		}%
	}
	\end{scope}
	\begin{scope}[xshift=\GW+\GX]
	\draw[rounded corners=3pt, thick, opacity=0.4]
	(-0.3cm, -0.3cm) rectangle (\GW+0.3cm, \GH+1.1cm);
	
	\node[font=\bfseries\large] at (.5*\GW, \GH + 0.6cm) {$P^2$};
	
	\foreach \r in {0,...,2}{%
		\foreach \c in {0,...,3}{%
			\pgfmathsetlengthmacro{\XX}{\c*(\CW+\PAD)}
			\pgfmathsetlengthmacro{\YY}{\r*(\CH+\PAD)}
			\coordinate (P2-r\r c\c) at (\XX+0.5*\CW, \YY);
			\coordinate (P2-r1\r c1\c) at (\XX+0.5*\CW, \YY-\CH);
			\ifnum \r = 0
		\ifnum \c = 2
		\path[draw, fill=ForestGreen!15] (\XX,\YY) rectangle ++(\CW,\CH); 
		\else
		\path[draw, fill=gray!5] (\XX,\YY) rectangle ++(\CW,\CH); 
		\fi
		\else
		\path[draw, fill=gray!5] (\XX,\YY) rectangle ++(\CW,\CH); 
		\fi
			\pgfmathsetlengthmacro{\XC}{\XX + 0.5*\CW}
			\pgfmathsetlengthmacro{\YC}{\YY + 0.5*\CH}
			\pgfmathrandominteger{\a}{1}{3}
			\pgfmathrandominteger{\b}{1}{4}
			\node at (\XC,\YC) {(\a,\b)};
		}%
	}%
	\end{scope}
	
	\begin{scope}[yshift=-(\GH+\GY)]
	\draw[rounded corners=3pt, thick, opacity=0.4]
	(-0.3cm, -0.3cm) rectangle (\GW+0.3cm, \GH+1.1cm);
	
	\node[font=\bfseries\large] at (.5*\GW, \GH + 0.6cm) {$P^3$};
	
	\foreach \r in {0,...,2}{%
		\foreach \c in {0,...,3}{%
			\pgfmathsetlengthmacro{\XX}{\c*(\CW+\PAD)}
			\pgfmathsetlengthmacro{\YY}{\r*(\CH+\PAD)}
			\coordinate (P3-r\r c\c) at (\XX+0.5*\CW, \YY+\CH);
			\coordinate (P3-r1\r c1\c) at (\XX+0.5*\CW, \YY+\CH + 0.5*\CH);
			\ifnum \r = 2
		\ifnum \c = 1
		\path[draw, fill=ForestGreen!15] (\XX,\YY) rectangle ++(\CW,\CH); 
		\else
		\path[draw, fill=gray!5] (\XX,\YY) rectangle ++(\CW,\CH); 
		\fi
		\else
		\path[draw, fill=gray!5] (\XX,\YY) rectangle ++(\CW,\CH); 
		\fi
			\pgfmathsetlengthmacro{\XC}{\XX + 0.5*\CW}
			\pgfmathsetlengthmacro{\YC}{\YY + 0.5*\CH}
			\pgfmathrandominteger{\a}{1}{3}
			\pgfmathrandominteger{\b}{1}{4}
			\node at (\XC,\YC) {(\a,\b)};
		}%
	}%
	\end{scope}
	\begin{scope}[xshift=\GW+\GX, yshift=-(\GH+\GY)]
	\draw[rounded corners=3pt, thick, opacity=0.4]
	(-0.3cm, -0.3cm) rectangle (\GW+0.3cm, \GH+1.1cm);
	
	\node[font=\bfseries\large] at (.5*\GW, \GH + 0.6cm) {$P^4$};
	
	\foreach \r in {0,...,2}{%
		\foreach \c in {0,...,3}{%
			\pgfmathsetlengthmacro{\XX}{\c*(\CW+\PAD)}
			\pgfmathsetlengthmacro{\YY}{\r*(\CH+\PAD)}
			\coordinate (P4-r\r c\c) at (\XX+0.5*\CW, \YY+\CH);
			\coordinate (P4-r1\r c1\c) at (\XX+0.5*\CW, \YY+\CH+0.5*\CH);
				\ifnum \r = 2
			\ifnum \c = 0
			\path[draw, fill=ForestGreen!15] (\XX,\YY) rectangle ++(\CW,\CH); 
			\else
			\path[draw, fill=gray!5] (\XX,\YY) rectangle ++(\CW,\CH); 
			\fi
			\else
			\path[draw, fill=gray!5] (\XX,\YY) rectangle ++(\CW,\CH); 
			\fi
			\pgfmathsetlengthmacro{\XC}{\XX + 0.5*\CW}
			\pgfmathsetlengthmacro{\YC}{\YY + 0.5*\CH}
			\pgfmathrandominteger{\a}{1}{3}
			\pgfmathrandominteger{\b}{1}{4}
			\node at (\XC,\YC) {(\a,\b)};
		}%
	}%
	\end{scope}
	
\draw[red, dashed, thick, -] (P4-r2c0) -- (P4-r12c10);	
\draw[red, dashed, thick, -] (P4-r12c10) -- (P3-r12c11);
\draw[red, dashed, thick, ->] (P3-r12c11) -- (P3-r2c1);
\draw[red, dashed, thick, ->] (P3-r2c1) -- (P2-r0c2);
\draw[red, dashed, thick, -] (P2-r0c2) -- (P2-r10c12);
\draw[red, dashed, thick, -] (P2-r10c12) -- (P1-r20c23);
\draw[red, dashed, thick, -] (P1-r20c23) -- (P1-r12c13);
\draw[red, dashed, thick, ->] (P1-r12c13) -- (P1-r2c3);
\end{tikzpicture}

%% file: proof.tex
\section{Proof of the Main Result}\label{sec:full-proof}

In this section, we prove \Cref{res:main} in its entirety. We employ the impossibility of the pointer chasing problem stated in \Cref{prop:pc-lb}. We start with a protocol for pointer chasing that uses an algorithm that approximately estimates the set cover. We describe the protocol in three sequential parts. Finally, we prove the correctness of our protocol, and use \Cref{prop:pc-lb} to prove the required space lower bound for set cover.

\subsection{Protocol for Pointer Chasing}

In this section, we see how to create a protocol $\prot$ for $\PC{p, a, b}$ using an algorithm $\alg$ for $\alpha$-approximation of set cover in $p$ passes over a universe $\cU$ of size $n$ and totally $m$ sets. 
The parameters $a$ and $b$ are chosen based on $m, n, \alpha$ and $p$ as follows.

\paragraph{Parameters:}
\begin{equation}\label{eq:params}
	a = \frac{m-1}{p} \qquad\textnormal{ and }\qquad b = \frac1{p+1} \cdot \paren{\frac{n}{10(p+1)\log m}}^{1/p} \cdot \frac1{\alpha^{(p+1)/p}}.
\end{equation}

Our reduction crucially uses public randomness to create the sets that the set cover algorithm is run on, and does not use any private randomness.
We break the protocol into three simple parts. 

We want to create $a$ sets for each player $\player{i}$ for $i \in [p]$ so that each of the $a$ sets corresponds to $b$ many function outputs held by $\player{i}$. Since only one function output of $\player{p+1}$ matters, $\player{p+1}$ only creates one set. 
For $\player{i}$ with $i \in [p]$, the $a \cdot b$ function values are grouped together in the natural way: $\player{i}$ creates one set $\Sset{i}{j}$ for each $j \in [a]$ that depends only on $\func{i}(j,k)$ for all $k \in [b]$. To continue, we first create many auxiliary sets called the \textbf{U-common family} with public randomness as follows. 

\subsubsection*{Construction of the U-common family:}

The first part of the protocol for pointer chasing involves the creation of many auxiliary sets using public randomness, and we describe this next. 
\begin{tbox}
	\textbf{Part $(1)$ Protocol $\prot$ for $\PC{p, a, b}$:}  \\
	Each player $\player{i}$ for $i \in [p]$ samples the following using \underline{public randomness}: 
	\begin{enumerate}[label=$(\roman*)$]
		\item For each $j \in [a]$, sample a set $\Uset{i}{j}$ uniformly at random of size $n/(10\alpha \cdot (p+1))$ from $[n]$, independently of all the other sets. 
		\item Randomly partition $\Uset{i}{j}$ into $b$ sets, each of size $n/(10\alpha \cdot b \cdot (p+1))$ using public randomness, called $\Uset{i}{j, k}$ for $ k \in [b]$. 
	\end{enumerate}
	Refer to \Cref{fig:Uset} for an illustration. 
\end{tbox}

\bigskip

\begin{figure}[h!]
	\begin{center}
		\input{tikz-Uset}
	\end{center}
	\caption{ Creation of U-common family for each player: this is repeated for all $\player{1}, \player{2}, \player{3}, \ldots, \player{p}$. The illustration is when $a = 3$ and $b = 4$. There are three sets $U_1, U_2$ and $U_3$ each of size $n/(10\alpha \cdot (p+1))$ chosen uniformly at random. These are split into four sets, each of size $n/(10\alpha \cdot b \cdot (p+1)) = n/(40\alpha \cdot (p+1))$. Although the sets $U_1, U_2$ and $U_3$ are shown as disjoint, they may have common elements as they are sampled independently. However, for each $i \in [3]$, the sets $U_{i,j}$ for $j \in [4]$ are disjoint from each other. }
	\label{fig:Uset}
\end{figure}

We will prove some simple properties of the sets in the U-common family. Mainly, we want to look at the intersection of any $p$ sets with one set each picked from the first $p$ players.   Although the below statement applies to the sets corresponding to the special pointers in \Cref{def:pc}, it holds for any collection of $p$ sets with the aforementioned property.

\begin{claim}\label{clm:last-universe-size}
	The set $\bigcap_{i \in [p]} \Uset{i}{z_i}$ has size at least $ 4 \alpha \cdot  (p+1) \cdot \log (m)$ with high probability over the randomness of the choices of all the sets $\Uset{i}{j, k}$ for $i \in [p+1], j \in [a], k \in [b]$. 
\end{claim}
\begin{proof}
	Each element is chosen to be in set $\Uset{i}{z_i}$ with probablity exactly $1/(\alpha \cdot b \cdot (p+1))$ as the set contains $n/(\alpha \cdot b \cdot (p+1))$ elements chosen uniformly at random. Moreover, the choices of $\Uset{i}{z_i}$ and $\Uset{i'}{z_{i'}}$ for $i \neq i'$ are independent of each other. 
	
	Hence, for any element $e$, the probability that it belongs to $\Uset{i}{z_i}$ for all $i \in [p]$ is:
	\begin{align*}
		\Pr[ e \in \cap_{i \in [p]} \Uset{i}{z_i}] = \paren{\frac1{\alpha \cdot b \cdot (p+1)}}^{p} = \frac1{\alpha^{p}} \cdot \paren{\frac{10 (p+1)\log m}{n}} \cdot \alpha^{p+1} = \frac1{n} \cdot (10 \alpha \cdot (p+1) \cdot \log m),
	\end{align*}
	by the value of $b$ from \Cref{eq:params}.
	
	Let $T$ denote the intersection of all the sets $\Uset{i}{z_i}$ for $i \in [p]$.
	For any element $e \in \cU$, let $X_e$ denote the indicator random variable for whether element $e$ is in $T$. We know that the $X_e$ random variables are negatively correlated, as $\Uset{i}{z_i}$ consists of exactly $n/(10\alpha \cdot b \cdot (p+1))$ elements. 
	Using \Cref{prop:chernoff}, we get, 
	\begin{align*}
		\Pr[\card{T} < 4 \alpha (p+1) \cdot \log m] &= \Pr[\sum_e X_e < 4  \alpha (p+1) \cdot (\log m)] \\
		&= \Pr[\sum_e X_e < (4/10) \Exp[\sum_e X_e]] \tag{by linearity of expectation} \\
		&\leq \exp(-0.18 \cdot 10 \alpha \cdot (p+1) \log m) \tag{by \Cref{prop:chernoff}} \\
		&\leq m^{-\alpha \cdot (p+1)},
	\end{align*}
	which proves the claim, as $\alpha \geq 1$ and for large $m$. 
\end{proof}

\begin{figure}[h!]
	\centering
	\subcaptionbox{The sets for $\player{1}$.}{
		\input{tikz-Sset-P1}
	}
	
	\bigskip
	
	\subcaptionbox{The sets for $\player{2}$.}{
		\input{tikz-Sset-P2}
	}
	
	\caption{An illustration of the S-private family for players $\player{1}$ and $\player{2}$ with $a = 3, b = 4$. Each $S^i_j$ for $i \in [2], j \in [3]$ is a union of all the sets to its right. }
	\label{fig:Sset}
\end{figure}

\subsubsection*{Creation of Set Cover Instance:} 
The next part of the protocol is to create the sets that the algorithm $\alg$ is run on. Each players only use the U-common sets and their own pointers in the pointer chasing instance to create the set cover instance. This step is performed deterministically by all the players. The sets will be collectively called the \textbf{S-private family}.

\begin{tbox}
	\textbf{Part $(2)$ Protocol $\prot$ for $\PC{p, a, b}$:}  \\
	\textbf{Input:} The U-common sets from Part $(1)$ of $\prot$, and $\PC{p,a,b}$ instance.  \\
	\textbf{Output:} Set cover instance. 
	\begin{enumerate}[label=$(\roman*)$]
	\item The last player $\player{p+1}$ constructs one input set $\Sset{p+1}{}$ as follows:
\[
\Sset{p+1}{} :=[n] \setminus \Uset{p}{\func{p+1}(1,1)}.
\]
\item Each other player $\player{i}$ for $1 < i \leq p$ constructs $a$ sets for $j \in [a]$ as follows:
\[
\Sset{i}{j} := \bigcup_{k \in [b]} \Uset{i}{j,k} \setminus \Uset{i-1}{\func{i}(j,k)}.
\]
\item The first player $\player{1}$ constructs $a$ input sets $\Sset{1}{j}$ for $j \in [a]$ as follows:
\[
\Sset{1}{j} : = \bigcup_{\substack{k \in [b]; \\ \func{1}(j,k) = 1}} \Uset{1}{j,k}.
\]
\end{enumerate}
Refer to \Cref{fig:Sset} for an illustration. 
\end{tbox}

Again, we prove some simple properties about the set cover instance. 

\begin{observation}\label{obs:num-of-sets}
	There are exactly $m$ sets created by all the $p+1$ players together.
\end{observation}
\begin{proof}
	Each player $\player{i}$ for $i \in [p]$ creates $a$ sets and the last player $\player{p+1}$ creates one set. Together, these are $a \cdot p + 1 = m$ sets by \Cref{eq:params}.
\end{proof}

The main lemma of this subsection follows. This states a gap in the values of the optimal set cover based on the answer of the pointer chasing instance. 

\begin{lemma}\label{lem:main-set-system}
	When the answer to the pointer chasing instance is $1$, there are $(p+1)$ sets among $m$ sets that cover the entire universe $[n]$. Otherwise, with high probability, any collection of $\alpha \cdot (p+1)$ sets among the $m$ sets does not cover the universe $[n]$. 
\end{lemma}
Before we prove \Cref{lem:main-set-system}, we establish some notation. 
For $i \in [p]$, we use $\Sstar{i}$ to denote the set $\Sset{i}{j}$ such that $z_i = (j, k)$ for some $k \in [b]$. These are the sets in the S-private family that depend on the special pointers given to the first $p$ players. We also use $\Sstar{p+1}$ to denote the set $\Sset{p+1}{}$. 

First, let us see what happens when the answer to the pointer chasing instance is $1$. The following fairly direct claim states that the size of the set cover instance is small, and this happens just by picking all $\Sstar{i}$ for $i \in[p+1]$. 

\begin{claim}\label{clm:pc-1}
	When the answer to the pointer chasing problem is $1$, the set cover instance constructed in protocol $\prot$, has a set cover of size at most $(p+1)$. 
\end{claim}
\begin{proof}
	We know that the answer to the pointer chasing instance is $\func{1}(z_1)$. 
	We will show that :
	\[
	\bigcup_{i \in [p+1]} \Sstar{i} = [n],
	\]
	which proves the claim.
	
We have that $\Sstar{p+1} = [n] \setminus \Uset{p}{z_{p}}$ by construction so we need to show that $\Uset{p}{z_p}$ is covered by the remaining sets. 
	We will show by induction on $i$ for $i \in [p]$, that when $\func{1}(z_1) = 1$, the sets $\Sstar{1},\Sstar{2}, \ldots, \Sstar{i}$ cover all the elements in $\Uset{i}{z_i}$, which will complete the proof. This will ensure that all elements in $\Uset{p}{z_p}$ will be covered by the union of $\Sstar{1}, \ldots, \Sstar{p}$ when $i = p$. 
	
	For the base case, we know directly that $\Sstar{1} = \Uset{1}{z_1}$, as $\func{1}(z_1) = 1$. Assume towards the induction hypothesis that $\Sstar{1}, \ldots,  \Sstar{i-1}$ covers all of $\Uset{i-1}{z_{i-1}}$. Then, the set $\Sstar{i}$ contains as a subset the set $\Uset{i}{z_i} \setminus \Uset{i-1}{z_{i-1}}$, and hence the union of $\Sstar{1}, \ldots, \Sstar{i}$ covers all the remaining elements of the set $\Uset{i}{z_i}$ by construction. 
\end{proof}

Now, we analyze what happens when the answer to the pointer chasing instance is $0$. The intersection of all $\Uset{i}{z_i} $ for $i \in [p]$, which are the sets corersponding to special pointers, will be large by \Cref{clm:last-universe-size}, and we will show that less than $\alpha \cdot (p+1)$ sets from the S-private family will not be able to cover all these elements.

\begin{claim}\label{clm:pc-0}
	When the answer to pointer chasing is $0$, with high probability no collection of $\alpha \cdot (p+1) $ sets $T_1, T_2, \ldots, T_{\alpha \cdot (p+1)}$ among the $m$ sets in S-private family covers the entire universe $[n]$. 
\end{claim}
\begin{proof}
	We will show specifically that the elements of $\bigcap_{i \in [p]} \Uset{i}{z_i}$ will not be covered by any collection of $\alpha \cdot (p+1)$ sets from the S-private family. 
	Let $T$ be the set $\bigcap_{i \in [p]} \Uset{i}{z_i}$. For each $i \in [p+1]$, let $\jstar_i \in [a]$ be the index such that $z_i = (\jstar_i, k)$ for some $k \in [b]$, i.e., $\jstar$ corresponds to the first part of the special pointer.
	Notice also that $\Sstar{i} = \Sset{i}{\jstar_i}$ for each $i \in [p]$. 
	
	We argue that $\Sstar{1}$ does not contain any element of set $T$ first using the fact that $T \subseteq \Uset{1}{z_1}$. Let $z_1 = (\jstar_1, k_1)$, and observe that $\Uset{1}{\jstar, k'}$ is disjoint from $\Uset{1}{\jstar_1, k_1}$ for any $k' \neq k_1, k' \in [b]$ by construction. Moreover, the set $\Uset{1}{z_1} = \Uset{1}{\jstar_1, k_1}$ is explicitly excluded from $\Sset{1}{\jstar_1}$ as the answer to the pointer chasing instance is 0. Hence, $\Sstar{1} = \Sset{1}{\jstar_1}$ is a subset of $\Uset{1}{\jstar_1} \setminus \Uset{1}{z_1}$, and does not contain any element of $\Uset{1}{z_1}$. 
	Similarly, we can argue that $\Sstar{i}$ for any $i \in [p]$ does not contain any element of $T$ as $T \subset \Uset{i}{z_i}$. 
	Again, as $T \subset \Uset{p}{z_{p}}$, the set $\Sstar{p+1}$ does not cover any element of $T$ either. 
	
	It remains to argue about any collection of $\alpha \cdot (p+1)$ sets from the S-private family excluding all $\Sstar{i}$ for $i \in [p+1]$. 
	To do this, we condition on the event that the set $T$ has at least $4\alpha (p+1) \cdot \log m$ elements. This event happens with high probability by \Cref{clm:last-universe-size}. For the rest of this proof, we do not use any randomness associated with any $\Uset{i}{\jstar_i}$ for any $i \in [p]$. This randomness is only used to condition on the size of the set $T$. 
	
	We will prove the stronger statement that any collection of $\alpha \cdot (p+1)$ sets $\Uset{i}{j}$ for $i \in [p], j \in [a] \setminus \set{\jstar_i}$ does not cover $T$, as $\Sset{i}{j} \subseteq \Uset{i}{j}$ for all $i \in [p], j \in [a]$. 
	We know that each $\Uset{i}{j}$ is a set of $n/10\alpha \cdot (p+1)$ elements chosen at random, independently of each other and of $T$. We will use Chernoff bounds to argue that the probability that they cover the entirety of set $T$ is small. 
	Fix any choice of $\alpha (p+1)$ distinct sets $\Uset{i}{j}$ for $i \in [p], j \in [a] \setminus \set{\jstar_i}$. 
	
	Choose any $4 \alpha (p+1)  \cdot (\log m)$ elements of $T$. These many distinct elements exist in $T$ because of our earlier conditioning.
	Let $X_{e, i}$ denote the indicator random variable for whether $e$ is covered by the $i^{\textnormal{th}}$ set for some $e$ in the subset of $4\alpha(p+1) \cdot (\log m)$ elements of $T$ and, $i \in [\alpha \cdot (p+1)]$.
	We have that:
	\begin{align*}
		\Exp[X_{e, i}] = \frac{1}{10\alpha \cdot (p+1)}, 
	\end{align*}
	as the $i^{\textnormal{th}}$ set chooses $n/(10\alpha \cdot (p+1))$ elements at random independently of $T$. The expectation of the sum of these random variables is:
	\begin{align}\label{eq:inter-1}
		\Exp[X] = \Exp[\sum_{e, i} X_{e, i}] = 4 \alpha (p+1)\cdot \log m\cdot \alpha \cdot (p+1) \cdot  \frac{1}{10\alpha(p+1)} = 4\alpha \cdot (\log m) \cdot (p+1)/10,
	\end{align}
	by linearity of expectation.
	We also have, 
	\begin{align*}
		&\Pr[\textnormal{$T$ is covered by the collection}] \\
		&\leq \Pr[ X\geq 4\alpha (p+1) \cdot \log m] \tag{as this subset of elements of $T$ must be covered } \\
		&\leq \Pr[X > 10 \cdot \Exp[X]] \tag{by \Cref{eq:inter-1}} \\
		&\leq \exp(-2\alpha \cdot (p+1) \cdot(\log m)). \tag{by \Cref{prop:chernoff}, and as the variables are negatively correlated}
	\end{align*}
	
	There are at most $\binom{m}{\alpha(p+1)}$ many $\alpha(p+1)$-size collection from the $m$ sets we are considering. A union bound over all such collections gives that with probability at most $m^{-\alpha(p+1)}$, the set $T$ is covered by some $\alpha$-size collection. For large $m$ and $\alpha \geq 1$, this probability is small, and thus the proof of the claim is complete. 
\end{proof}

The proof of \Cref{lem:main-set-system} now follows directly from \Cref{clm:pc-0} and \Cref{clm:pc-1}.

\subsubsection*{Running the streaming algorithm $\alg$:} 
The final part in the construction of $\prot$ is to actually run $\alg$ and output an answer. This part is straightforward.

\begin{tbox}
	\textbf{Part $(3)$ of protocol $\prot$:} \\
	\textbf{Input:} Pointer Chasing instance $\PC{p, a, b}$ and a $p$-pass $\alpha$-approximation algorithm $\alg$ for set cover over a universe  $[n]$, taking $m$ input sets. \\
	 \textbf{Output}: Answer to pointer chasing instance.
	\begin{enumerate}[label=$(\roman*)$]
		\item The players run $\alg$ on the $m = a\cdot (p-1) + 1$ input sets $\Sset{p}{}$ and $\Sset{i}{j}$ for $i \in [p], j \in [a]$ over the universe $[n]$ by sending messages to the blackboard. (See \Cref{clm:running-alg} for how to run $\alg$.)
		\item If $\alg $ returns a value $\leq \alpha \cdot (p+1)  $ as an estimate of the size of the optimal set cover, $\player{p}$ outputs that the value of the pointer chasing instance is 1. Otherwise, $\player{p}$ outputs 0.
	\end{enumerate}
\end{tbox}

\noindent

First, we show that the players can run $\alg$ effectively.
\begin{claim}\label{clm:running-alg}
	Algorithm $\alg$ that uses $p$-passes and $s$ space can be run by the $(p+1)$ players using total communcation $(p+1)\cdot p \cdot s$ total communication and $p$ rounds. 
\end{claim}

\begin{proof}
	Using the sets $\Uset{i}{j, k}$ sampled with public randomness for $i \in [p], j \in [a], k \in [b]$, we have:
	\begin{itemize}
		\item $\player{i}$ knows all the sets $\Sset{i}{j}$ for $i \in [p], j \in [a]$ as they depend only on the publicly chosen U-common family, and the pointers given to $\player{i}$. 
		\item $\player{p+1}$ knows the set $\Sset{p+1}{}$ as it depends on the publicly sampled U-common family, and the value of $\func{p}(1,1)$, that $\player{p}$ knows.
	\end{itemize}
	Hence, collectively the players can run one pass of $\alg $ starting from $\player{1}$ who inserts the sets $\Sset{1}{j}$ for $j \in [a]$, and sends the memory state of $\alg$ to the blackboard in $s$ bits. This is followed by $\player{2}$ who inserts $\Sset{2}{j}$ for all $j \in [a]$, all the way till $\player{p+1}$ who inserts $\Sset{p+1}{}$. The total communication used to run one pass of $\alg$ is $(p+1) \cdot s$ bits, and can be done in one round. 
	
	To run all the $p$-passes, $p \cdot (p+1) \cdot s$ bits of messages are used totally. 
\end{proof}

Next, we argue the correctness of protocol $\prot$.

\begin{claim} \label{clm:prot-correct}
	If algorithm $\alg$ succeeds with probability at least $2/3$, protocol $\prot $ for $\PC{p, a, b}$ succeeds with probability at least $3/5$. 
\end{claim}
\begin{proof}
	When the answer to pointer chasing is 1, by \Cref{lem:main-set-system}, we know that the optimal set cover is of size at most $(p+1)$. Hence, if $\alg$ is correct, it outputs an estimate of size at most $\alpha \cdot (p+1)$, and $\prot$ outputs the correct answer as well. 
	When the answer is $0$, by \Cref{lem:main-set-system}, we know that with high probability, the set cover size is more than $\alpha \cdot (p+1)$. Thus, if $\alg$ succeeds, the estimate of the size of the set cover will be more than $ \alpha \cdot (p+1)$. Protocol $\prot$ succeeds with probability at least $3/5$ for large values of $m,n$ and $\alpha \geq 1$.
\end{proof}

This concludes our description of the protocol. As we have also proved the necessary properties along the way, proving \Cref{res:main} now is simple, and is covered in the next subsection.

\subsection{Finishing the Proof of \Cref{res:main}}

 We can complete the proof of \Cref{res:main} using the impossibility result for pointer chasing stated in \Cref{prop:pc-lb}, and our proof of the correctness of protocol $\prot$ in \Cref{clm:prot-correct}.

\begin{proof}[Proof of \Cref{res:main}]
	Firstly, let us examine if parameters $a, b$ as specified in \Cref{eq:params} can be chosen. 
	We have $p \ll m$, so we can choose value of $a$ as required. We also have $p \cdot \alpha = o(n/\log m)^{1/(p+1)}$, so parameter $b$ can also be set to the required large value. Thus, pointer chasing instances of the input size specified in \Cref{eq:params} can be constructed. 
	
	We know algorithm $\alg$ uses $p$-passes and $s$-space to provide an $\alpha$-approximate estimate of the solution to a set cover instance over universe $[n]$ with $m$ sets, and succeed with probability at least $2/3$.
	We know that protocol $\prot$ uses $(p+1)p\cdot s$ bits of communication from \Cref{clm:running-alg}, and from \Cref{clm:prot-correct}, we know it succeeds with probability at least $ 3/5$ for large $m$ and $\alpha \geq 1$. 
	
	Therefore, by \Cref{prop:pc-lb} and choice of parameters in \Cref{eq:params}, we know that:
	\begin{align*}
		s \geq \frac1{(p+1)p} \cdot \Omega(\frac{ab}{p^5}) = \Omega(1/p^7) \cdot \Omega\paren{\frac{(m-1)}{p} \cdot \paren{\frac{n}{\alpha }}^{1/(p)} \cdot \frac{1}{\alpha(p+1)} \cdot \frac{1}{(10\log m \cdot (p+1))^{1/p}}}.
	\end{align*}
	By simplifying the above term, we know that $\alg$ uses space at least:
	\[
		\Omega\paren{\frac{1}{p^{10} \cdot \log m} \cdot \frac{m}{\alpha} \cdot  \paren{\frac{n}{\alpha}}^{1/p}},
	\]
	which finishes the proof.
\end{proof}

%% file: tikz-Uset.tex
\usetikzlibrary{calc,decorations.pathreplacing,positioning}

\begin{tikzpicture}[font=\small]
	\def\SetW{12cm}          
	\def\SetH{1.2em}        
	\def\VGap{1.5cm}        
	
	\pgfmathsetlengthmacro{\PartW}{\SetW/4}
	
	\foreach \i in {1,2,3} {
		\begin{scope}[yshift={-(\i-1)*\VGap}]
			\draw[line width=1pt, rounded corners=2pt]
			(0,0) rectangle (\SetW,\SetH);
			
			\foreach \k in {1,2,3} {
				\draw[line width=0.7pt]
				(\k*\PartW,0) -- (\k*\PartW,\SetH);
			}
			
			\node[font=\normalsize, yshift=1.0em] at (\SetW/2,\SetH) {$U_{\i}$};
			
			\foreach \j in {1,...,4} {
				\node at ({(\j-0.5)*\PartW}, {0.5*\SetH}) {$U_{\i,\j}$};
			}
			
			\ifnum\i=3
			\draw [decorate,decoration={brace,mirror,amplitude=1.5pt}]
			(0,-1.0em) -- (\PartW,-1.0em)
			node[midway,below=3pt] {$\displaystyle n/(10\alpha(p+1) \cdot b)$};
			
			\draw [decorate,decoration={brace,mirror,amplitude=6pt}]
			(0,-2.5em) -- (\SetW,-2.5em)
			node[midway,below=6pt] {$\displaystyle n/(10\alpha(p+1))$};
			\fi
		\end{scope}
	}
\end{tikzpicture}

%% file: tikz-Sset-P1.tex

\begin{tikzpicture}[x=1cm,y=1cm,>=stealth]
	\def\CWL{1.40} 
	\def\CWR{1.5} 
	\def\CH{0.85}  
	\def\PAD{0.14} 
	\def\PADL{0.14}
	\def\GAP{1.8}  
	
	\pgfmathsetmacro{\TW}{\CWR-0.20} 
	
	\coordinate (L0) at (0,0); 
	\coordinate (R0) at ($ (L0) + ({4*\CWL + 3*\PADL + \GAP}, {0}) $); 
	
	\newcommand{\setbit}[3]{\expandafter\gdef\csname bit-#1-#2\endcsname{#3}}
	\newcommand{\getbit}[2]{%
		\ifcsname bit-#1-#2\endcsname
		\csname bit-#1-#2\endcsname
		\else
		0%
		\fi
	}
	
	\setbit{1}{1}{0}\setbit{1}{2}{1}\setbit{1}{3}{0}\setbit{1}{4}{1}
	\setbit{2}{1}{1}\setbit{2}{2}{0}\setbit{2}{3}{1}\setbit{2}{4}{0}
	\setbit{3}{1}{0}\setbit{3}{2}{0}\setbit{3}{3}{1}\setbit{3}{4}{1}
	
	\node[font=\bfseries] at ($ (L0) + ({0.5*(4*\CWL + 3*\PADL)}, {0.6}) $) {$P^1$};
	
	\foreach \i in {1,2,3}{
		\foreach \j in {1,2,3,4}{
			\coordinate (cL-\i-\j) at ($ (L0) + ({(\j-1)*(\CWL+\PADL)}, {-(\i-1)*(\CH+\PAD)}) $);
			\draw[rounded corners=1pt, line width=0.5pt]
			(cL-\i-\j) rectangle ++(\CWL,-\CH);
			\node[font=\small] at ($ (cL-\i-\j) + ({\CWL/2}, {-\CH/2}) $) {\getbit{\i}{\j}};
		}
	}
	
	\foreach \i in {1,2,3}{
		\node[font=\bfseries] at ($ (R0) + ({-0.70}, {-(\i-1)*(\CH+\PAD) - 0.5*\CH}) $) {$S^{1}_{\i}$};
		
		\foreach \j in {1,2,3,4}{
			\coordinate (cR-\i-\j) at ($ (R0) + ({(\j-1)*\CWR}, {-(\i-1)*(\CH+\PAD)}) $);
			\fill[gray!10] (cR-\i-\j) rectangle ++(\CWR,-\CH);
			\draw[line width=0.6pt] (cR-\i-\j) rectangle ++(\CWR,-\CH);
			
			\edef\valnum{\getbit{\i}{\j}}%
			\ifnum\valnum=0\relax
			\node[font=\scriptsize, align=center, text width=\TW cm]
			at ($ (cR-\i-\j) + ({\CWR/2}, {-\CH/2}) $)
			{$\emptyset$};
			\else
			\node[font=\scriptsize, align=center, text width=\TW cm]
			at ($ (cR-\i-\j) + ({\CWR/2}, {-\CH/2}) $)
			{$U^{1}_{\i,\j}$};
			\fi
		}
	}
\end{tikzpicture}

%% file: tikz-Sset-P2.tex

\begin{tikzpicture}[x=1cm,y=1cm,>=stealth]
	\def\CWL{1.40} 
	\def\CWR{2.0} 
	\def\CH{0.85}  
	\def\PAD{0.14} 
	\def\PADL{0.14}
	\def\GAP{1.8}  
	
	\pgfmathsetmacro{\TW}{\CWR-0.20} 
	
	\coordinate (L0) at (0,0); 
	\coordinate (R0) at ($ (L0) + ({4*\CWL + 3*\PADL + \GAP}, {0}) $); 
	
	\newcommand{\setpair}[4]{%
		\expandafter\gdef\csname a-#1-#2\endcsname{#3}%
		\expandafter\gdef\csname b-#1-#2\endcsname{#4}%
	}
	\newcommand{\geta}[2]{%
		\ifcsname a-#1-#2\endcsname \csname a-#1-#2\endcsname \else 1\fi
	}
	\newcommand{\getb}[2]{%
		\ifcsname b-#1-#2\endcsname \csname b-#1-#2\endcsname \else 1\fi
	}
	
	\setpair{1}{1}{1}{2}\setpair{1}{2}{3}{4}\setpair{1}{3}{2}{1}\setpair{1}{4}{2}{3}
	\setpair{2}{1}{3}{1}\setpair{2}{2}{1}{4}\setpair{2}{3}{2}{2}\setpair{2}{4}{3}{3}
	\setpair{3}{1}{2}{4}\setpair{3}{2}{3}{2}\setpair{3}{3}{1}{1}\setpair{3}{4}{1}{3}
	
	\node[font=\bfseries] at ($ (L0) + ({0.5*(4*\CWL + 3*\PADL)}, {0.6}) $) {$P^2$};
	
	\foreach \i in {1,2,3}{
		\foreach \j in {1,2,3,4}{
			\coordinate (cL-\i-\j) at ($ (L0) + ({(\j-1)*(\CWL+\PADL)}, {-(\i-1)*(\CH+\PAD)}) $);
			\draw[rounded corners=1pt, line width=0.5pt]
			(cL-\i-\j) rectangle ++(\CWL,-\CH);
			\node[font=\scriptsize] at ($ (cL-\i-\j) + ({\CWL/2}, {-\CH/2}) $)
			{$\big(\geta{\i}{\j},\getb{\i}{\j}\big)$};
		}
	}
	
	\foreach \i in {1,2,3}{
		\node[font=\bfseries] at ($ (R0) + ({-0.70}, {-(\i-1)*(\CH+\PAD) - 0.5*\CH}) $) {$S^{2}_{\i}$};
		
		\foreach \j in {1,2,3,4}{
			\coordinate (cR-\i-\j) at ($ (R0) + ({(\j-1)*\CWR}, {-(\i-1)*(\CH+\PAD)}) $);
			\fill[gray!10] (cR-\i-\j) rectangle ++(\CWR,-\CH);
			\draw[line width=0.6pt] (cR-\i-\j) rectangle ++(\CWR,-\CH);
			
			\edef\x{\geta{\i}{\j}}%
			\edef\y{\getb{\i}{\j}}%
			
			\node[font=\scriptsize, align=center, text width=\TW cm]
			at ($ (cR-\i-\j) + ({\CWR/2}, {-\CH/2}) $)
			{$U^{2}_{\i,\j}\setminus U^{1}_{\x,\y}$};
		}
	}
\end{tikzpicture}

%% file: figfull.tex
\begin{figure}[p!]
	
	\vspace{-0.75cm}
	\centering
	\hspace{-1cm}
	\subcaptionbox{Three main regime of parameters studied previously. We note that the low approximation regime is a box
		in the tradeoff domain whereas the other two are planes.\label{fig:tradeoff1}}%
	[.48\linewidth]{
		\includegraphics[scale=0.38]{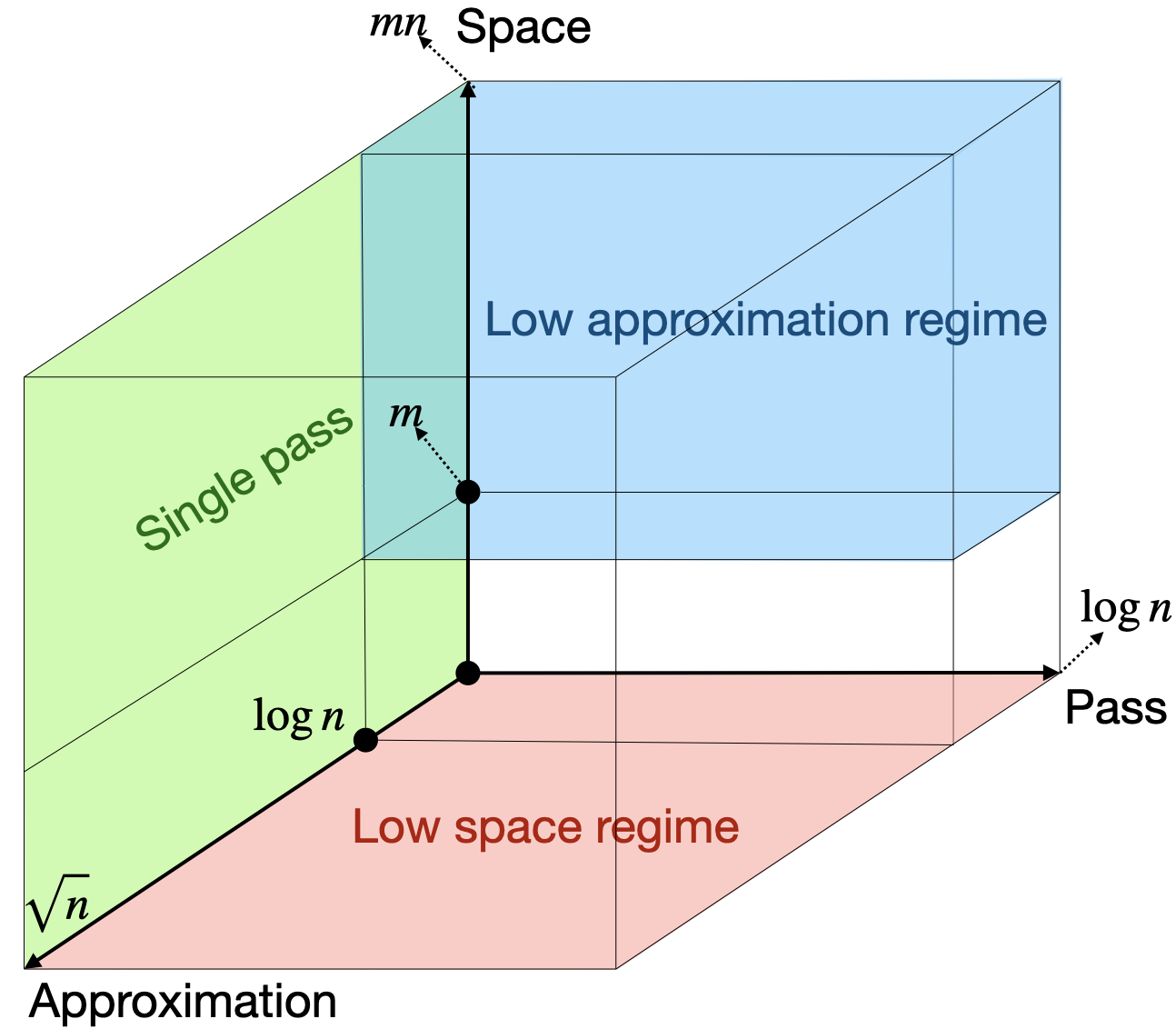}
	}
	\hspace{1.25cm}
	\subcaptionbox{Main prior lower bound curves. Each of these lower bounds 
		is fixed in one dimension of the tradeoff. The shaded area denotes the range ruled out by the lower bound. The curves for single-pass and low-space regime are 
		optimal but the low-approximation curve is not. 
		\label{fig:tradeoff2}}%
	[.48\linewidth]{
		\includegraphics[scale=0.38]{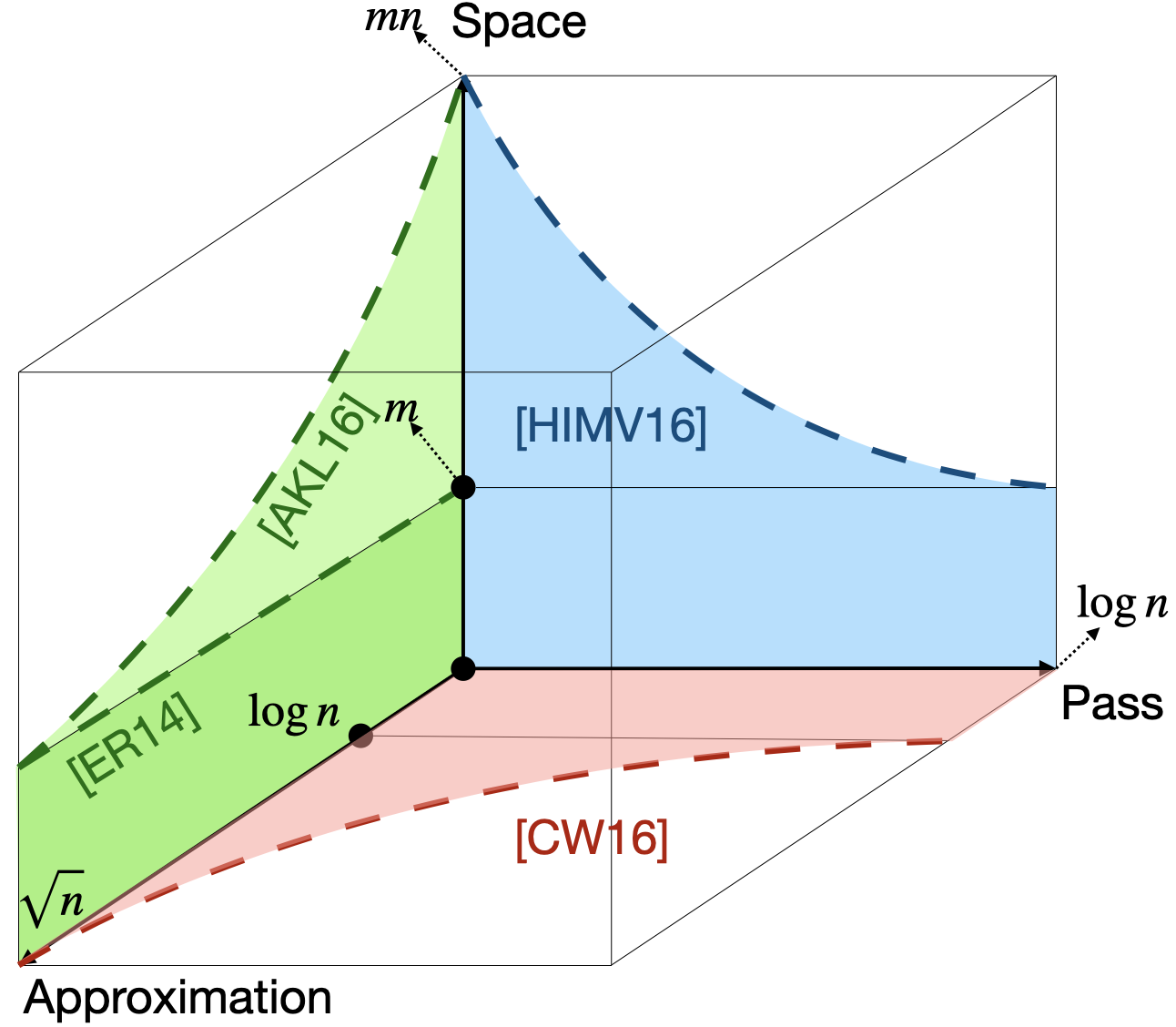}
	}
	
	\vspace{0.5cm}
	\hspace{-1cm}
	\subcaptionbox{Two other lower bound curves in the low approximation regime when approximation $<\log{n}$. The curves subsume the previous lower bound of~\cite{HarPeledIMV16} from the last part
		and are again optimal. The tradeoff plane for existing algorithms are also drawn: for $\log{n}$-approximation (and beyond), no space-pass lower bounds were known when space is $\Omega(m)$. 
		\label{fig:tradeoff3}}%
	[.47\linewidth]{
		\includegraphics[scale=0.38]{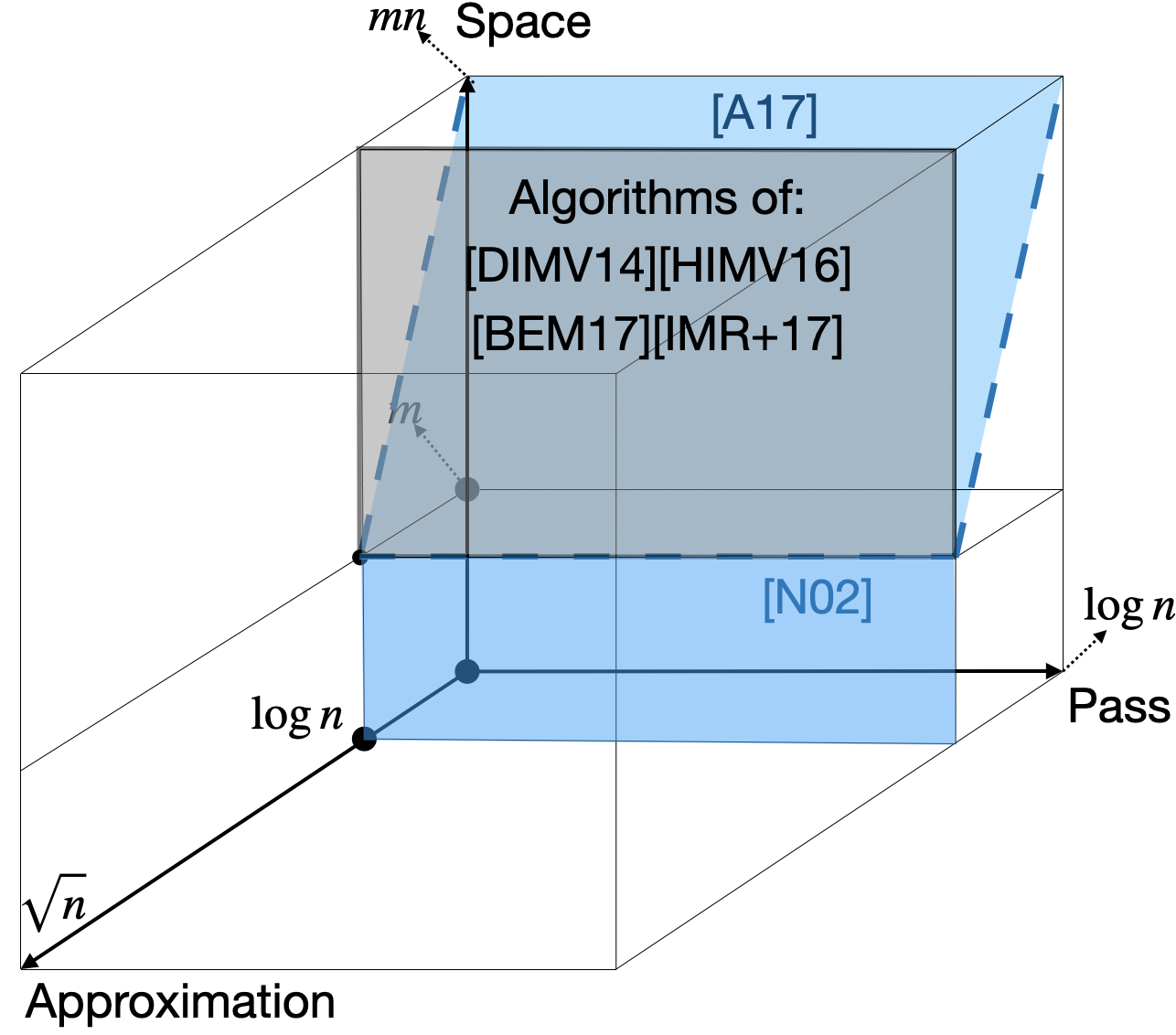}
	}
	\hspace{1.25cm}
	\subcaptionbox{Our lower bound addresses the entire regime of approximation at least $\log{n}$ and is optimal throughout. 
		Projecting our tradeoff curve to the single-pass plane recovers the lower bound of~\cite{AssadiKL16} and to the low-space plane recovers the lower bound of~\cite{ChakrabartiW16} (up to logarithmic factors). 
		The \emph{most interesting} projection is to the low-space plane (at $\log{n}$ approximation) which had no prior lower bounds; the projection matches the performance of existing algorithms on this plane also.
		\label{fig:tradeoff4}}%
	[.47\linewidth]{
		\includegraphics[scale=0.38]{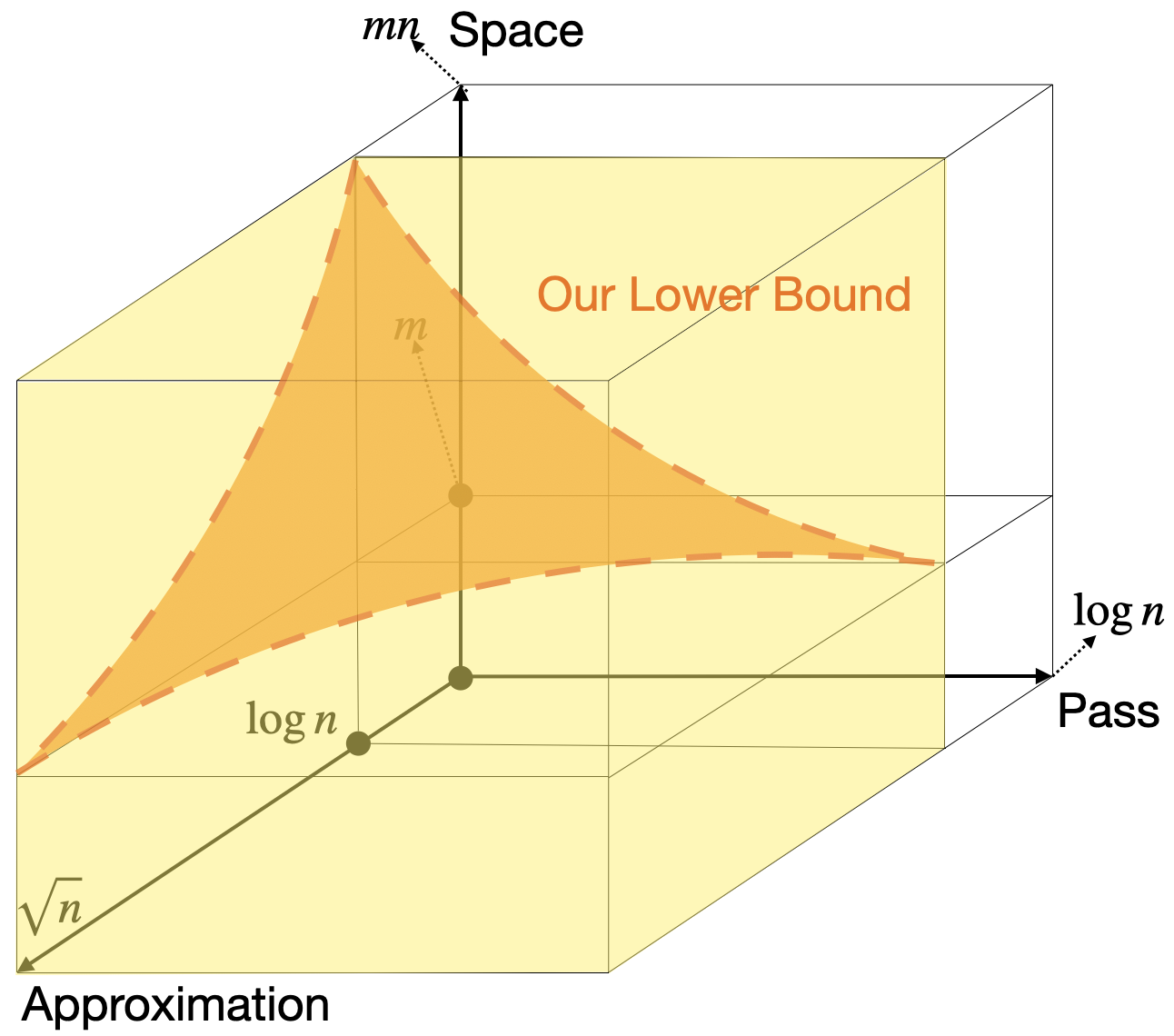}
	}
	\caption{An illustration of the three-way tradeoff of \emph{space vs pass vs approximation} for streaming set cover. The origin of the axes is at ($n$ space, $1$ pass, and $1$ approximation); the drawn curves are for illustrative purpose and do not necessarily reflect the precise tradeoff curve. }
	\label{fig:tradeoff}
\end{figure}

%% file: thealg.tex
\section{An Algorithm for Streaming Set Cover Estimation}
\label{sec:setcov-ub}

\newcommand{\probsamp}{\ensuremath{q}}
\newcommand{\cUsamp}{\ensuremath{\cU'}}
\newcommand{\optset}{\ensuremath{\textnormal{opt}}}
\newcommand{\cK}{\ensuremath{\mathcal{K}}}
\newcommand{\otil}{\ensuremath{\hat{o}}}
\newcommand{\probcov}{\ensuremath{r}}
\newcommand{\optfam}{\ensuremath{\mathcal{O}}}
\newcommand{\cA}{\ensuremath{\mathcal{A}}}
\newcommand{\cUest}{\ensuremath{\cU_{\textnormal{samp}}}}
\newcommand{\cUleft}{\ensuremath{\cU_{\textnormal{left}}}}
\newcommand{\algo}{\ensuremath{\textnormal{ALG}}}
\newcommand{\cSstar}{\ensuremath{\cS^{\star}}}
\newcommand{\sol}{\ensuremath{\cS^{\star}}}
\newcommand{\nest}{\ensuremath{n_{\textnormal{samp}}}}

\newcommand{\cUpruned}{\ensuremath{\cU_{\textnormal{pr}}}}
\newcommand{\solpruned}{\ensuremath{\sol_{\textnormal{pr}}}}

\newcommand{\kstar}{\ensuremath{k^{\star}}}

For completeness, we now give an algorithm that matches our tradeoff in \Cref{res:main} up to poly-logarithmic factors in $n, m$ and constant factors in $p$.
This result is a simple modification of the algorithms in \cite{AssadiKL16,HarPeledIMV16} for set cover estimation in one pass and finding an approximate set cover in multiple passes, respectively.

\begin{proposition}[A combination of~\cite{AssadiKL16,HarPeledIMV16}]\label{thm:setcov-ub}
For any $n, m, p \in \IN$ and $\alpha \geq 1$, there exists an $O(p)$-pass streaming algorithm that, given a family of $m$ sets over a universe of size $n$, finds an $\alpha$-approximate set cover in \[
	\Ot(\frac{m}{\alpha} \cdot \paren{\frac{n}{\alpha}}^{1/p}),
\]
space with high probability. 
\end{proposition}

We use some primitives from \cite{AssadiKL16} and \cite{HarPeledIMV16}, but we prove all of them for completeness. We need some notation before we proceed. 

\paragraph{Notation.}
Given a universe $\cU$ and a family $\cS$ of sets $S_1, \ldots, S_m$, for any $\cK\subseteq \cU$, we use $\optfam(\cK) \subseteq \cS$ to denote any optimal set cover from the family $\cS$ that covers $\cK$ (this need not cover the entire universe), and $\optset(\cK)$  to denote the size of this set cover. 

\subsection{Estimating Set Cover Size in One Pass}
\cite{AssadiKL16} gives an algorithm that estimates the size of an optimal set cover up to an $\beta$-factor in a single pass in $\Ot(mn/\beta^2)$ space. The main tool they use is the following element sampling lemma.

\begin{lemma}[\!\!\cite{AssadiKL16}]\label{lem:subsample-est}
 Let $\cU$ be a universe with $n$ elements and $\cUsamp$ be a subset of $\cU$ obtained by sampling each element of $\cU$ independently with probability \[
	\probsamp:= \frac{4 \ln n}{\beta}.
	\]
	Then,
	\[
	\Pr\paren{\optset(\cUsamp) < \optset(\cU) \cdot \frac1{8\beta}} \leq \frac{1}{4}.
	\]
\end{lemma}
\begin{proof}
	Define 
	\[
	t := {\frac{q}{32\ln{n}}} \cdot \optset(\cU)+1 = \optset(\cU) \cdot \frac1{8 \beta} + 1.
	\]
	To prove this lemma we need the following claim, which shows that the inclusion of any one fixed element in the smaller universe $\cUsamp$ cannot increase the size of its optimal set cover by a large margin. 
	\begin{claim}\label{clm:remove-element}
		For a sampled smaller universe $\cUsamp$ as in the statement of~\Cref{lem:subsample-est} and any element $e\in \cU$, we have
		\[
		\Pr\Paren{\optset(\cUsamp)< t \mid e\in \cUsamp} \geq \Pr\Paren{\optset(\cUsamp) < t - 1}.
		\]
	\end{claim}
	\begin{proof}
		Consider $\cUsamp \setminus \{e\}$, the smaller universe obtained by removing $e$ from $\cUsamp$ (if it exists). Then,
		\begin{align*}
			\Pr\Paren{\optset(\cUsamp)< t  \mid e \in \cUsamp} &\geq \Pr\Paren{\optset(\cUsamp \setminus \{e\}) < t-1 \mid e\in \cUsamp}\\
			&= \Pr\Paren{\optset(\cUsamp \setminus \set{e}) < t-1};
		\end{align*}
		where the first inequality holds because, conditioned on $e \in \cUsamp$, if $\optset(\cUsamp \setminus \set{e}) < t-1$, we can find a set cover of $\cUsamp$ using at most $t$ sets by including one additional set that covers $e$. The next equality holds because $\cUsamp \setminus \set{e}$ is a random subset of $\cU \setminus \set{e}$, obtained by picking each element independently with probability $q$. 
		Finally, as $\optset(\cUsamp \setminus \set{e}) \leq \optset(\cUsamp)$, we have, 
		\[
	\Pr\Paren{\optset(\cUsamp \setminus \set{e}) < t-1} \geq \Pr\Paren{\optset(\cUsamp) < t - 1}.
		\]
		proving the claim. \Qed{clm:remove-element} 
		
	\end{proof}
	We now continue the proof of~\Cref{lem:subsample-est}. 
	Assume towards a contradiction that,
	\begin{align}\label{eq:prob-sample-contradiction}
		\Pr\paren{\optset(\cUsamp) < \optset(\cU) \cdot \frac1{8\beta}}=\Pr\Paren{\optset(\cUsamp) < t - 1} > \frac{1}{4}.
	\end{align}      
	Let $\mu$ be the distribution of the smaller universe $\cUsamp$ \emph{conditioned} on the event that $\optset(\cUsamp)< t$ (we emphasize that this is $t$ and not $t-1$ unlike in~\Cref{eq:prob-sample-contradiction}). 
	We sample $k:=4 \beta$ smaller universes $\cUsamp_i$ for $i \in [k]$, each independently from $\mu$. 
	For any element $e \in \cU$ and $i \in [k]$, we have,
	\begin{align*}
		\Pr_{\mu}\Paren{e \in \cUsamp_i} &= \Pr\Paren{e \in \cUsamp \mid \optset(\cUsamp) < t},
		\end{align*}
		where in the RHS, we sample smaller universe $\cUsamp_i$ exactly as in the lemma statement. The equality above holds by the definition of $\mu$. We continue as:
	\begin{align*}
		 \Pr\Paren{e \in \cUsamp \mid \optset(\cUsamp) < t} &=\frac{\Pr \Paren{e \in \cUsamp} \cdot \Pr\Paren{\optset(\cUsamp) < t\mid e \in \cUsamp}}{\Pr\Paren{\optset(\cUsamp)< t}} \tag{by Bayes' rule}\\
		&\geq\frac{\Pr \Paren{e \in \cUsamp} \cdot \Pr\Paren{\optset(\cUsamp) < t-1}}{\Pr\Paren{\optset(\cUsamp)< t}}  \tag{by~\Cref{clm:remove-element}}\\
		&> \frac{q}{4} \cdot \frac{1}{\Pr\Paren{\optset(\cUsamp)< t}} \tag{by~\Cref{eq:prob-sample-contradiction} and process of sampling $\cUsamp$} \\
		&\geq \frac{\ln n}{\beta}. \tag{by value of $q$ and the denominator is at most one}
	\end{align*}
	So in sampling of $\cUsamp_1,\ldots,\cUsamp_k $ independently from $\mu$, the probability that element $e \in \cU$ is not sampled in any of the smaller universes is,
	\begin{align*}
		\Pr\Paren{e \notin \cUsamp_i \textnormal{ for all $i \in [k]$}}\leq \paren{1-\frac{\ln n}{\beta}}^{k} \leq \exp(-\frac{\ln n}{\beta} \cdot (4\beta)) = \frac1{n^4}.
	\end{align*}
	By a union bound over all the elements of $\cU$, we get that with high probability, there exists a collection of smaller universes $\cUsamp_1, \ldots, \cUsamp_k$ for $k = 4\beta$ such that,
	\begin{enumerate}[label=$(\roman*)$]
	\item Each $\cUsamp_i$ for $i \in [k]$ has $\optset(\cUsamp_i) <t = \optset(\cU)/8\beta+1$. 
	\item Each element $e \in \cU$ is present in at least one of $\cUsamp_i$ for $i \in [k]$.
	\end{enumerate} 
	Thus, we can cover the entirety of $\cU$ by taking a union of the optimal covers of $\optset(\cUsamp_i)$ for $i \in [k]$. This has cost at most $(\optset(\cU)/8\beta+1) \cdot 4\beta = \optset(\cU)/2+4\beta < \optset(\cU)$, which is a contradiction 
	(note that we can assume $4\beta < \optset(\cU)/2$ as otherwise the probability-term in the statement of the lemma will be $\Pr\paren{\optset(\cUsamp) < 1}$ which is trivially $0$). 
	This shows that \Cref{eq:prob-sample-contradiction} cannot hold, proving the lemma statement. 
\end{proof}

For any guess $\otil$ of $\optset(\cU)$, \Cref{lem:subsample-est} directly gives way to check if $\otil$ is an appropriate estimate of $\optset(\cU)$. Sample $\cUsamp$ as in the statement of \Cref{lem:subsample-est}, and check $\optset(\cUsamp) \leq \otil$. Accept $\otil$ if so, and reject otherwise. Naively implementing this algorithm takes $\Ot(mn/\beta)$-space but we show that this space requirement can be improved with multiple passes using the primitive of \cite{HarPeledIMV16}.

\subsection{Covering Original Universe with Smaller Samples}

\Cref{lem:subsample-est} shows that sampling a smaller universe can give an accurate estimate of the size of the optimal set cover. \cite{HarPeledIMV16} proves that covering the smaller randomly sampled universe also covers a large fraction of the original universe with high probability. 

\begin{lemma}[\!\!\cite{HarPeledIMV16}]\label{lem:subsample-cover}
	 Let $\cU$ be a universe with at most $n$ elements and $\cUsamp$ be a subset of $\cU$ obtained by sampling each element of $\cU$ independently with probability \[
	\probcov:= 100 \cdot \optset(\cU) \cdot \log m \cdot \log n \cdot n^{\gamma-1},
	\]
	for some $\gamma \in (0,1)$.  Then, with probability at least $(1-1/n^2)$, any set family $\optfam(\cUsamp)$ that covers $\cUsamp$ leaves at most $n^{1-\gamma}$ elements of $\cU$ as uncovered. 
\end{lemma}

\begin{proof}
	Let $\opt(\cU)$ be $\ell$, and let $\cA \subseteq \cS$ be any family of at most $\ell$ sets from $\cS$ that do not cover at least $n^{1-\gamma}$ elements of $\cU$. 
	If $\cA$ is chosen as a family that is the optimal set cover of $\cUsamp$, then, none of the elements that $\cA$ does not cover are sampled as a part of $\cUsamp$. Therefore, 
	\begin{align*}
		\Pr\paren{\textnormal{$\cA$ covers $\cUsamp$}} &\leq (1-r)^{n^{1-\gamma}} \tag{none of the elements not covered by $\cA$ are sampled} \\
		&\leq \exp(-r \cdot n^{1-\gamma}) 
		= \exp(-100 \cdot \ell \cdot \log m \cdot \log n \cdot n^{\gamma-1} \cdot n^{1-\gamma})\leq m^{-4 \ell} \cdot n^{-2}. 
	\end{align*}
	There are at most $m^{2\ell}$ possible choices for $\cA$ by a trivial bound over the number of subsets of $\cS$ that contain at most $\ell$ elements. Hence, by a union bound over all possible choices of $\cA$, we get that with probability at least $1-n^{-2}$, such a set $\cA$ that leaves uncovered $n^{1-\gamma}$ elements of $\cU$ does \emph{not} end up as a set cover of $\cUsamp$. Here, we have also used that $\optset(\cUsamp) \leq \optset(\cU) = \ell$, as we only do a union bound over all subsets of $\cS$ of size at most $\ell = \optset(\cU)$.  
\end{proof}

With these two components, we are ready to give the final algorithm. 

\subsection{Estimating Set Cover Size}
We use \Cref{lem:subsample-est} to estimate set cover size. However, we do not find an optimal set cover of the sample $\cU'$ but instead find an approximate set cover of $\cUsamp$ over multiple passes using \Cref{lem:subsample-cover}. The algorithm is given an integer $k$ and it decides whether $\opt(\cU) \geq k$ or $\opt(\cU) = o(k/\alpha)$, i.e., whether or not $k$ is a $\Theta(\alpha)$-approximation to  $\opt(\cU)$. 

We sample a set $\cUsamp$ such that we can find an $(\alpha/p)$-approximation to the set cover of $\cU$ using $\cUsamp$, similar to \Cref{lem:subsample-est}. Then, we use $O(p)$ passes to find a $O(p)$-approximate set cover of $\cUsamp$ in low space.

\begin{tbox}
\textbf{Algorithm $\algo$ for checking size of set cover.}

\textbf{Input}: value $k$, input family $\cS$, universe $\cU$, number of iterations $p$, and approximation factor $\alpha$.
\begin{enumerate}[label=$(\roman*)$]
\item\label{line:step-i}  Let $\beta = \alpha/p$. Sample a smaller universe $\cUest$ where each element in $\cU$ is independently chosen with probability $q:= (4 \ln n)/\beta$, as in \Cref{lem:subsample-est}. We have the following definitions:
\begin{align*}
		\nest &:= \card{\cUest}; \tag{the number of elements in the sample} \\
	\cUleft&:=\cUest; \tag{the set of uncovered elements} \\
	\cUpruned &:= \emptyset; \tag{the set of elements which will be covered by pruning} \\
	\solpruned &:= \emptyset; \tag{the sets which will cover elements of $\cUpruned$} \\
	\sol &:= \emptyset. \tag{the sets which will cover the remaining elements of $\cUest$} 
\end{align*} 
\item Repeat the following for $i = 1, \ldots, p$ iterations:
\begin{enumerate}
	\item \label{line:bundle-1}From $\cUleft$, sample $\cU_i \subseteq \cUleft$, where each element of $\cUleft$ is chosen independently with probability $r:= 100 \cdot k \cdot \log m \cdot \log n \cdot (\nest)^{(i/p) - 1}$. Let $\cS_i = \emptyset$, and $n_i$ be the size of $\cU_i$ after this sampling step.
	\item\label{line:bundle-2} In one pass, for $j \in [m]$, for each incoming set $S_j$, do the following: 
	
	(Pruning step) If $\card{S_j \cap \cU_i} > n_i/(k\beta)$, add $S_j$ to $\solpruned$, add elements of $S_j$ to $\cUpruned$ and update $\cU_i$ as $\cU_i \setminus S_{j}$. 
	
	(Storing step) Otherwise, store $S_j \cap \cU_i$ explicitly, and add $S_j$ to family $\cS_i$. 
	\item\label{line:reject} Find an optimal set cover $\cSstar_i$ of $\cU_i$ using sets from the family $\cS_i$, as $\cS_j \cap \cU_i$ is stored in memory for all $S_j \in \cS_i$. If the size of this set cover $\cSstar_i$ is more than $k$, \textbf{reject} the input value $k$.
	\item\label{line:collect} Otherwise, update $\cUleft$ as $\cUleft \setminus \big( \paren{\cup_{S \in \cSstar_i} S} \cup \cUpruned \big)$. Add all the sets in $\cSstar_i$ to $\sol$. 
\end{enumerate}
\item \textbf{Accept} value of $k$ if none of the previous iterations reject. 
\end{enumerate}
\end{tbox}

We show that the above algorithm, with high probability, accepts $k$ if it is at least $\optset(\cU)$, and rejects $k$ if it is is at most $\optset(\cU)/32\alpha$. Running the algorithm $O(\log m)$ times in parallel for geometrically increasing guesses of $k$ gives an algorithm for correctly estimating the set cover value. 

We begin by arguing that when the value of $k$ is large enough, the algorithm always accepts. 
\begin{claim}\label{clm:alg-accept-k}
For any $k \geq \optset(\cU)$, $\algo$ always accepts. 
\end{claim}
\begin{proof}
$\algo$ only rejects is in Line~\eqref{line:reject} of the algorithm, and only if the sampled universe has an optimal set cover of size strictly larger than $k$. 
Since $\cU_i \subseteq \cU$ always, we have $\opt(\cU_i) \leq \opt(\cU) \leq k$ and thus Line~\eqref{line:reject} never rejects in this case. 
\end{proof}

Now, we show that if value of $k$ is too low, the algorithm rejects $k$. We will find a set cover of $\cU$ parametrized by $k$ whenever the algorithm accepts, which will prove that $k$ is not too low.

\begin{claim}\label{clm:prune-step}
When $\algo$ accepts, there is a set cover of $\cUpruned$ of size at most $\beta \cdot k \cdot p$. 
\end{claim}
\begin{proof}
	In any iteration $i \in [p]$, the total number of sets added to $\solpruned$ in the pruning step in Line~\eqref{line:bundle-2} is at most $\beta\cdot k$. 
	For any fixed $i \in [p]$, every time any set $S_j$ is added to $\solpruned$, the size of $\cU_i$ decreases by at least an $n_i/(k\beta)$ additive factor, where  $n_i$ was the size of set $\cU_i$ before any sets are processed. Therefore,  the pruning step can happen at most $\beta \cdot k$ times before the set $\cU_i$ becomes empty.  Over all $p$ iterations, at most $\beta \cdot k \cdot p$ sets are added to $\solpruned$. 
	
	 The only elements added to $\cUpruned$ are elements of some set $S_j$ that is also added to $\solpruned$ in some iteration. Therefore, $\solpruned$ forms a set cover of $\cUpruned$ trivially.
\end{proof}

We then show that the remainder of $\cUest$ can also be covered cheaply when $\algo$ accepts, based on the samples we pick in each iteration. 

\begin{claim}\label{clm:size-bound-Uleft}
With probability at least $1-1/n$, for any $0 \leq i \leq p$, after iteration $i$, the size of $\cUleft $ is at most $(\nest)^{1-(i/p)}$.
\end{claim} 
\begin{proof}
	The proof follows by an induction on $p$. When $p = 0$, the size of $\cUsamp$ is $\nest$ by definition. We use \Cref{lem:subsample-cover} to argue that the size reduces after each iteration. 
	
	For the induction hypothesis, we assume that the size of $\cUleft$ is at most $(\nest)^{1-i/p}$ after iteration $i$. Consequently $\cUleft \setminus \cUpruned$ is also of size at most $(\nest)^{1-i/p}$.
	In the next $(i+1)$-iteration, each element in $\cUleft \setminus \cUpruned$ is chosen independently with probability \[
	r = 100 \cdot k \cdot \log m \cdot \log n \cdot (\nest)^{(i/p)-1} \cdot (\nest)^{1/p}.
	\]
The set family $\cSstar_i$ leaves at most $(\nest)^{1-(i/p)} \cdot (\nest)^{-1/p}$ elements of $\cUleft \setminus \cUpruned$ uncovered by \Cref{lem:subsample-cover}. The size of $\cUleft\setminus \cUpruned$ decreases to at most $(\nest)^{1-(i+1)/p}$ after this iteration with probability at least $1-n^{-2}$. As elements of $\cUpruned $ are automatically removed from $\cUleft$ in Line~\eqref{line:collect}, so this size bound holds for $\cUleft$ at the end of iteration $(i+1)$.

	With a naive union bound that \Cref{lem:subsample-cover} holds over all iterations, and as $p \leq n$, we prove the claim. 
\end{proof}

We know now that the size of $\cUleft$ decreases after each iteration. It is not hard to prove that the sets we pick also cover the entirety of $\cUleft \setminus \cUpruned$. 

\begin{observation}\label{clm:bad-case-opt}
	When $\algo$ accepts, with probability at least $(1-1/n)$, $\opt(\cUest \setminus \cUpruned) \leq k \cdot p$.
\end{observation}
\begin{proof}
	We will show that $\sol$ is a set cover of $\cUest \setminus \cUpruned$. In each iteration, at most $k$ sets are added to $\sol$. Thus totally after $p$ iterations, at most $k \cdot p$ sets would be added. Proving that $\sol$ covers all the elements of $\cUest \setminus \cUpruned$ would be sufficient. 
By \Cref{clm:size-bound-Uleft}, after the last $p^{\textnormal{th}}$ iteration, the size of $\cUleft$ is empty with probability at least $1-1/n$. 	
We start with $\cUleft = \cUest$, and whenever we remove an element from $(\cUleft \setminus \cUpruned)$, it is covered by some set in $\sol$. Thus, at the end of the last iteration, $\cUleft$ will have no elements, and $\sol$ will be a valid set cover. 
\end{proof}

Finally, we are ready to show that when $\algo$ accepts, value of $k$ is not too small. 

\begin{claim}\label{clm:correct-k-small}
When $k < \optset(\cU)/32\alpha$, with probability at least $1/2$, $\algo$ rejects.  
\end{claim}
\begin{proof}
	Firstly, we argue that $\optset(\cU \setminus \cUpruned)$ is large. This is because, by \Cref{clm:prune-step} and value of $k$, 
	\begin{equation}\label{eq:inter-alg-1}\optset(\cUpruned) \leq \beta \cdot p \cdot k   \leq \frac{\alpha}{p} \cdot p \cdot \optset(\cU) \cdot \frac{1}{32\alpha}  = \optset(\cU) \cdot \frac{1}{32}.
	\end{equation}
	
	We can cover $\cU$ by simply taking a set cover of $\cU\setminus \cUpruned$ and $\cUpruned$ separately. We also have, 
	\[
		\optset(\cU) \leq \optset(\cUpruned) + \optset(\cU \setminus \cUpruned),
	\]
	which from \Cref{eq:inter-alg-1}, gives that,
	\begin{equation}\label{eq:inter-alg-2}
		\optset(\cU \setminus \cUpruned) \geq \optset(\cU) \cdot \frac{31}{32} > \optset(\cU) \cdot \frac{3}{4}.
		\end{equation}
	Next, we argue that $\cUest \setminus \cUpruned$ is a subsample of $\cU \setminus \cUpruned$ where each element is chosen with probability $q = 4\ln n/\beta$: in Line~\ref{line:step-i} of $\algo$, the sampling process is independent, and each element is chosen with probability $q$ irrespective of the other elements. 
	
	By \Cref{lem:subsample-est} and using the value of $\beta$, we have that,
	\[
		\optset(\cUest \setminus \cUpruned) \geq \optset(\cU \setminus \cUpruned) \cdot \frac{1}{8\beta} = \optset(\cU \setminus \cUpruned) \cdot \frac{p}{8\alpha},
	\] 
	with probability at least $3/4$. Using \Cref{eq:inter-alg-2}, we also have,
	\begin{equation}\label{eq:inter-alg-3}
		\optset(\cUest \setminus \cUpruned) \geq \optset(\cU) \cdot \frac{3p}{32 \alpha}.
	\end{equation}
	On the other hand, from \Cref{clm:bad-case-opt}, we get,
	\[
		\optset(\cUest \setminus \cUpruned) \leq kp
	\]
	with probability at least $1-1/n$. With probability at least $3/4-(1/n) \geq (1/2)$, 
	\begin{align*}
			k \cdot p &\geq \optset(\cU) \cdot 3p \cdot \frac1{32\alpha}
	\end{align*}
	which shows that $k \geq \optset(\cU) \cdot (3/32\alpha)$, contradicting the bound on $k$ in the lemma statement. 
\end{proof}

\begin{claim}\label{clm:alg-space-usage}
	The space and passes used by $\algo$ are $\Ot(\frac{m}{\alpha} \cdot \paren{\frac{n}{\alpha}}^{1/p} + n)$ and $2p$ with high probability. 
\end{claim}

\begin{proof}
	We claim each iteration uses two passes which combined with having $p$ iterations bounds the number of passes. 
	In each iteration, one pass is used for  pruning followed by storing the sampled elements in  Lines~\eqref{line:bundle-1},~\eqref{line:bundle-2}, and~\eqref{line:reject}, and
	one more pass is used to find the elements covered by $\cSstar_i$, and remove them from $\cUleft$ in Line~\eqref{line:collect}. 
	
	Regarding space usage, all the sets $\cUest$, $\cUleft$ and $\cUpruned$ can be stored in $\Ot(n)$ space. The main requirement comes from Line~\eqref{line:bundle-2}, where $S_j \cap \cU_i$ is stored explicitly for every input set $S_j$ if it has at most $n_i/(k\beta)$ size. 
	The total space usage to store these elements in all the sets is at most:
	\[
		\tilde{O}\paren{\frac{mn_i}{k\beta}}.
	\]
	Giving an appropriate bound on $n_i$ is sufficient to complete the proof. 
	
	Using \Cref{clm:size-bound-Uleft}, we know that the size of $\cUleft$ at the beginning of iteration $i$ is at most $(\nest)^{1-(i-1)/p}$. As each element of $\cUleft$ is sampled independently, the expected number of elements in $\cU_i$ after Line~\eqref{line:bundle-1} is:
	\begin{align*}
		\Exp[\card{\cU_i}]& = \card{ \cUleft} \cdot r 
		\leq (\nest)^{1-(i-1)/p} \cdot 100k \log m \cdot \log n \cdot (\nest)^{i/p-1} = 	100 k \cdot \log m \cdot \log n \cdot (\nest)^{1/p}.
	\end{align*}
	As these elements are chosen independently with probability $r$, by an application of Chernoff bound (\Cref{prop:chernoff}), we can say that with probability at least $1-1/(mn^2)$, the value of $n_i$ which is the size of $\cU_i$ does not exceed twice its expectation.
	
	By a simple union bound over all sets and all iterations, we know that the number of elements stored totally in all the sets is at most 
	\[
		\Ot\paren{\frac{mn_i}{k\beta}} = \Ot( \frac{m}{k\beta} \cdot k \cdot (\nest)^{1/p}) = \Ot(\frac{m}{\alpha} \cdot(\nest)^{1/p}),
	\]
	wiith high probability, where we have used that $\beta = \alpha/p$, and $p \leq O(\log n)$.
	We will argue that $\nest \leq 8 \ln n \cdot n p /\alpha$,  with high probability as well, which will complete the proof of the claim. 
	
	Each element added to $\cUest$ is chosen with probability $4 \ln n /\beta = 4\ln n \cdot p/\alpha$, so in expectation, the size of this set is $4n \ln n \cdot p/\alpha$. Again by an application of \Cref{prop:chernoff}, we know that $\nest \leq 8 \ln n \cdot n \cdot p/\alpha$ with high probability. 	
\end{proof}


\begin{proof}[Proof of \Cref{thm:setcov-ub}]
	We run $\algo$ that checks if some input value $k$ is an appropriate estimate of $\optset(\cU)$ for $O(\log m)$ guesses of $k$ from $k = 2, 2^2, \ldots, 2^{\log m}$. For each guess of $k$, we run $\algo$ in parallel $O(\log n)$ times and accept $k$ if \emph{all} the parallel runs accept, and reject $k$ otherwise. By~\Cref{clm:correct-k-small,clm:alg-accept-k}, with high probability, $\algo$ accepts any $k \geq \optset(\cU)$, and rejects any $k < \optset(\cU)/32\alpha$. We return the lowest value of $k$ for which $\algo$ accepts as the estimate of $\optset(\cU)$. Let this value be $\kstar$ and thus have $\kstar \geq \frac1{32\alpha}\optset(\cU)$ with high probability.
Finally, we can re-parameterize $\alpha \leftarrow \alpha/32$ to obtain an $\alpha$-approximation as well without changing the space bounds asymptotically (if $\alpha < 32$, we find an $O(p)$-approximation in $p$-iterations directly from \cite{HarPeledIMV16}). This concludes the proof. 
\end{proof}